\documentclass[11pt]{article}

\usepackage{graphicx}
\usepackage{float}
\usepackage{chet}
\usepackage{scalerel}
\usepackage{amssymb}

\newcommand{\lsp}{\hspace{1pt}}
\newcommand{\llsp}{\hspace{0.5pt}}
\newcommand{\lnsp}{\hspace{-1pt}}

\newcommand{\veps}{\varepsilon}
\newcommand{\cO}{\mathcal{O}}
\newcommand{\cS}{\mathcal{S}}

\newcommand{\tetra}{{\hspace{-0.8pt}\triangle}}
\def\mcirc{\mathbin{\scalerel*{\bigcirc}{t}}}

\date{January 2018}

\preprint{CERN-TH-2018-012}

\title{Bootstrapping hypercubic and hypertetrahedral\\
\vspace{5pt}
theories in three dimensions}

\author{Andreas Stergiou}

\affiliation{Theoretical Physics Department, CERN, Geneva, Switzerland}

\abstract{There are three generalizations of the Platonic solids that exist
in all dimensions, namely the hypertetrahedron, the hypercube, and the
hyperoctahedron, with the latter two being dual. Conformal field theories
with the associated symmetry groups as global symmetries can be argued to
exist in $d=3$ spacetime dimensions if the $\veps=4-d$ expansion is valid
when $\veps\to1$.  In this paper hypercubic and hypertetrahedral theories
are studied with the non-perturbative numerical conformal bootstrap. In the
$N=3$ cubic case it is found that a bound with a kink is saturated by a
solution with properties that cannot be reconciled with the $\veps$
expansion of the cubic theory.  Possible implications for cubic magnets and
structural phase transitions are discussed. For the hypertetrahedral theory
evidence is found that the non-conformal window that is seen with the
$\veps$ expansion exists in $d=3$ as well, and a rough estimate of its
extent is given.}

\begin{document}

\maketitle

\newsec{Introduction}
The numerical conformal bootstrap~\cite{Rattazzi:2008pe} in three
dimensions has produced impressive results, especially concerning critical
theories in universality classes that also contain scalar conformal field
theories (CFTs). For the Ising universality class the bootstrap is in fact
the state-of-the-art method for the determination of critical
exponents~\cite{ElShowk:2012ht, El-Showk:2014dwa, Kos:2014bka,
Simmons-Duffin:2016wlq}. For theories with continuous global symmetries
there has also been significant progress, with important developments for
the Heisenberg universality class~\cite{Kos:2013tga, Kos:2015mba,
Kos:2016ysd}. In this work we study critical theories with discrete global
symmetries in three dimensions, focusing on the hypercubic and
hypertetrahedral symmetry groups.

The motivation for our considerations is twofold. First, in the cubic case
with $N=3$ scalar fields\foot{For $N=3$ the cubic and tetrahedral (with a
$\mathbb{Z}_2$) theories are equivalent.} the issue of its stability
relative to the $O(3)$ theory has, to our knowledge, remained unresolved;
see \cite{Pelissetto:2000ek, Tissier:2002zz} and references therein.
Second, in the hypertetrahedral case there exists a non-conformal window,
at least as far as the $\veps$ expansion is concerned~\cite{Zia:1975ha,
Osborn:2017ucf}, meaning that there is a range of the number $N$ of scalar
fields for which there are no hypertetrahedral fixed points, while such
fixed points can be found below and above that range. In this paper we
study these questions directly in $d=3$ with the bootstrap by considering a
single four-point function, namely that of the scalar operator $\phi_i,
i=1,\ldots,N$.

To our knowledge there have been no non-perturbative Monte Carlo studies of
hypercubic or hypertetrahedral CFTs in $d=3$, although the cubic
deformation at the $O(3)$ fixed point has been studied with Monte Carlo
methods in~\cite{Caselle:1997gf}. The standard lore (see
e.g.~\cite{Pelissetto:2000ek}) is thus mostly based on the perturbative
$\veps=4-d$ expansion~\cite{Wilson:1971dc} with the use of resummation
techniques~\cite{LeGuillou:1977rjt, LeGuillou:1979ixc}.  The $\veps$
expansion has proven to be quite powerful in the $O(N)$ and Ising models,
among others.  Nevertheless, it is hard to argue rigorously and in
generality about its effectiveness in $d=3$, or $\veps=1$. Therefore, it is
very important to check the results of the $\veps$ expansion whenever
non-perturbative methods are available.  In this paper we attempt to do
this for hypercubic and hypertetrahedral theories using the bootstrap. Our
expectation is that the predictions of the $\veps$ expansion should persist
beyond perturbation theory.  Various results for the hypercubic and
hypertetrahedral theories obtained with the $\veps$ expansion were recently
summarized in~\cite{Osborn:2017ucf}.  Many quantitative results for the
hypercubic theory can be found in~\cite{Pelissetto:2000ek} and references
therein, while some newer results can be found in~\cite{Dey:2016mcs},
notably pertaining to non-singlet and non-scalar operators.

One aim of this work is to provide general bounds on scaling dimensions of
operators in hypercubic and hypertetrahedral theories. In some cases these
bounds are found to display kinks. In the bootstrap we typically think of
kinks as special positions in parameter space where the bounds are
saturated by actual theories.  We are also interested in bounds on OPE
coefficients, especially the central charge. Such bounds also display
features that we attribute to saturation from actual CFTs.

In this work we find that the solution of crossing that saturates a certain
bound with a kink (Fig.~\ref{fig:Delta_V_cubic} below) has properties that
do not agree with the $\veps$ expansion. To be more specific, we are
referring to the $N=3$ case, where our bootstrap results show that the
first singlet scalar, i.e.\ what one would call the ``$\phi^2$'' operator,
has scaling dimension that differs significantly between the $O(3)$ fixed
point and the solution at the kink, called $C_3^b$ here. The $\veps$
expansion indicates that the ``$\phi^2$'' operator should have more or less
the same scaling dimension both at the $O(3)$ and the cubic theory
$C_3^\veps$ accessed with it~\cite{Aharony:1973zz, Carmona:1999rm,
Pelissetto:2000ek}.

Furthermore, when we study the stability of the $C_3^b$ solution, we find
another surprise. Based on the $\veps$ expansion results
of~\cite{Michel:1983in} it is expected that, if it exists, the stable fixed
point under a given set of deformations is unique.\foot{Note that this is
very special to the $\veps$ expansion in $4-\veps$ dimensions. For example,
it is not true in $6-\veps$ dimensions.} From our bootstrap analysis we
find that the $C_3^b$ solution is stable, i.e.\ it has only one relevant
scalar singlet operator (namely the ``$\phi^2$'' operator).  Now, stable
fixed points should describe physical systems at second order phase
transitions reached by tuning the temperature.  However, in cubic magnets,
where the cubic deformation is important, the critical exponents appear to
take the $O(3)$ or $C_3^\veps$ values according to
experiments~\cite{Pelissetto:2000ek}.  Since we find the critical exponents
of the $C_3^b$ solution to be very different from those of the  $C_3^\veps$
and $O(3)$ models, we see that despite the fact that the $C_3^b$ solution
is stable, it is the $O(3)$ or $C_3^\veps$ theory that describes cubic
magnets at criticality. From these observations one concludes that neither
the $C_3^b$ solution nor one of the $O(3)$ or $C_3^\veps$ theories has a
second relevant operator, directly contradicting the perturbative result
of~\cite{Michel:1983in}.  To our knowledge there is no contradiction with
having two stable fixed points non-perturbatively.

Our solution $C_3^b$ appears unrelated to the $C_3^\veps$ theory found with
the $\veps$ expansion.  We are unable to determine if the $C_3^b$ solution
at the kink corresponds to an actual CFT. Perhaps it does not, in which
case the kink is an artifact of the numerics. $C_3^b$ could also correspond
to a theory with cubic symmetry that cannot be obtained with the $\veps$
expansion. We are unable to settle this question in this work.

For the hypertetrahedral theories we find that our non-perturbative results
are consistent with the perturbative ones. Focusing on the non-conformal
window, we find that its range as estimated with the bootstrap is
reasonably close to that obtained with the $\veps$ expansion.  We should
note, however, that our bootstrap determination is approximate. We expect
that more accurate results can be obtained by considering a
mixed-correlator bootstrap.

The structure of the paper is as follows. In the next section we outline
some group theory that allows us to obtain the hypercubic crossing
equation. In section~\ref{boundsHyperC} we use the numerical bootstrap
method to find operator dimension bounds in various sectors of hypercubic
theories, analyze the spectrum of the solution at a bound, and discuss
extensively possible implications of our results both for cubic magnets and
for structural phase transitions. We also obtain bounds on the central
charge.  In section~\ref{hyperTetra} we derive the hypertetrahedral
crossing equation.  Finally, in section~\ref{boundsHyperT} we get operator
dimension bounds in hypertetrahedral theories and comment on the
non-conformal window. We conclude in section~\ref{conc}. In an appendix we
perform a bootstrap analysis of the cubic crossing equation in $d=3.8$.

\textbf{Note Added:} While this manuscript was in preparation
\cite{Rong:2017cow} appeared which also considered the bootstrap with
hypertetrahedral symmetry. There is a small overlap of their results with
ours in sections \ref{hyperTetra} and \ref{boundsHyperT}.

\newsec{Hypercubic symmetry}
For the hypercubic (or hyperoctahedral since hypercubes are dual to
hyperoctahedra) group at $N=4$ useful references
include~\cite{Baake:1981qe, Baake:1982ah}. For general $N$ a useful
reference is~\cite{Baake:1984uh}.

To start, let us go through the $N=3$ cubic symmetry group in some detail.
In Coxeter notation this is the group $C_3$ (or $B_3$).  It is given by the
semi-direct product $\cS_3\ltimes \mathbb{Z}_2{\lnsp}^3\simeq
\cS_4\ltimes\mathbb{Z}_2$, where $\cS_n$ is the permutation group of $n$
elements. $C_3$ is a subgroup of $O(3)$. The group $O(3)$ keeps the dot
product of two arbitrary three-vectors invariant.  If these vectors have
integer coefficients in the canonical $\mathbb{R}^3$ basis, then $C_3$ also
preserves that property.  This statement generalizes to
$C_N=S_N\ltimes\mathbb{Z}_2{\lnsp}^N$~\cite{Baake:1984uh}, which has
$2^NN!$ elements.

The irreducible representations of $C_3$ are those of $\cS_4$ (for each
parity due to the $\mathbb{Z}_2$), which are known to be the
$\boldsymbol{1}, \bar{\boldsymbol{1}}, \boldsymbol{2}, \boldsymbol{3}$ and
$\bar{\boldsymbol{3}}$. If $C_3$ is viewed as a subgroup of $O(3)$, it is
the traceless symmetric representation of $O(3)$ that gives rise to the
representations $\boldsymbol{2}$ and $\bar{\boldsymbol{3}}$. Terms that
belong to the diagonal make up the $\boldsymbol{2}$, while off-diagonal
terms give rise to the $\bar{\boldsymbol{3}}$. $\bar{\boldsymbol{1}}$ is
the one-dimensional sign representation, taking into account only the
signature of the permutation.  Finally, $\boldsymbol{1}$ is the trivial
representation and $\boldsymbol{3}$ corresponds to the antisymmetric
representation. For our purposes operators exchanged in the
$\phi_i\times\phi_j$ OPE need to be considered. There are singlets of even
spin and antisymmetric tensors of odd spin, and, instead of traceless
symmetric operators of even spin as in the $O(3)$ case, there are operators
of even spin in the $\boldsymbol{2}$ and $\bar{\boldsymbol{3}}$ of $C_3$.

It turns out that many statements in the previous paragraph generalize to
the $C_N$ case. For example, the singlet and antisymmetric representation
of $O(N)$ remain irreducible under $C_N$~\cite{Baake:1984uh}, while the
traceless-symmetric representation of $O(N)$ splits into two irreducible
representations under $C_N$, furnished separately by diagonal and
off-diagonal terms.

In the case of $O(N)$ symmetry the four-point function $\langle\phi_i(x_1)
\phi_j(x_2)\phi_k(x_3)\phi_l(x_4)\rangle$ was decomposed in conformal
blocks in~\cite{Rattazzi:2010yc, Kos:2013tga}. In the $12\to34$ channel,
for example,
\eqna{x_{12}^{2\Delta_\phi}x_{34}^{2\Delta_\phi}
\langle\phi_i(x_1)\phi_j(x_2)\phi_k(x_3)\phi_l(x_4)\rangle&=
\sum_{\text{S}^+_{\mcirc}}\lambda_\cO^2\lsp\delta_{ij}\delta_{kl}\lsp
g_{\Delta,\lsp\ell}(u,v)\\
&\quad+\sum_{\text{T}^+_{\mcirc}}\lambda_\cO^2\lsp\Big(\delta_{ik}
\delta_{jl}+\delta_{il}\delta_{jk}-\frac{2}{N}\delta_{ij}\delta_{kl}\Big)
\lsp g_{\Delta,\lsp\ell}(u,v)\\
&\quad+\sum_{\text{A}^{\lnsp-}_{\mcirc}}\lambda_\cO^2\lsp(\delta_{ik}
\delta_{jl}-\delta_{il}\delta_{jk})\lsp g_{\Delta,\lsp\ell}(u,v)\,,
}[ONConfBlock]
where three classes of operators contribute, namely even-spin singlets,
even-spin traceless-symmetric tensors, and odd-spin antisymmetric tensors,
as follows from the representation theory of $O(N)$.\foot{We use the
conventions of~\cite{Behan:2016dtz} for the conformal block
$g_{\Delta,\lsp\ell}(u,v)$.} In the hypercubic case the first and the last
term in \ONConfBlock remain the same, but the middle term gets further
decomposed under the hypercubic group. In particular, the diagonal terms,
with $i=j$ and $k=l$, need to be distinguished from the non-diagonal terms.

To proceed we introduce the tensors
\eqn{A_{ijkl}=\delta_{ijkl}\,,\qquad
B_{ijkl}=\delta_{ij}\delta_{kl}-\delta_{ijkl}\,,\qquad
\delta_{ijkl}=\begin{cases}
  1, & i=j=k=l\\
  0, & \text{otherwise}
\end{cases}\,.}[tensAB]
The tensor $B_{ijkl}$ is only symmetric under $i\leftrightarrow j$,
$k\leftrightarrow l$ and $ij\leftrightarrow kl$. This allows us to separate
the diagonal terms, with $i=j$ and $k=l$, from the non-diagonal terms in
\ONConfBlock.  Indeed, using \tensAB equation \ONConfBlock can be
decomposed to the hypercubic form
\eqna{x_{12}^{2\Delta_\phi}x_{34}^{2\Delta_\phi}
\langle\phi_i(x_1)\phi_j(x_2)\phi_k(x_3)\phi_l(x_4)\rangle&=
\sum_{\text{S}_\square^+}\,\lambda_\cO^2\lsp(A_{ijkl}+B_{ijkl})\lsp
g_{\Delta,\lsp\ell}(u,v)\\
&\quad+\sum_{\text{V}_\square^+}\lambda_\cO^2
\lsp\bigg(\Big(2-\frac{2}{N}\Big)A_{ijkl}-\frac{2}{N}B_{ijkl}\bigg)\lsp
g_{\Delta,\lsp\ell}(u,v)\\
&\quad+\sum_{\text{Y}_\square^+}\lambda_\cO^2\lsp
(B_{ikjl}+B_{iljk})\lsp g_{\Delta,\lsp\ell}(u,v)\\
&\quad+\sum_{\text{A}_\square^{\lnsp-}}\lambda_\cO^2
\lsp(B_{ikjl}-B_{iljk})\lsp g_{\Delta,\lsp\ell}(u,v)\,,
}[CubConfBlock]
where there are now four classes of operators that contribute, since the
even-spin $\text{T}^+_{\mcirc}$ operators of \ONConfBlock decompose into
the $\text{V}_\square^+$ and $\text{Y}_\square^+$ operators under
hypercubic symmetry. Equation~\eqref{CubConfBlock} has appeared already
in~\cite{Dey:2016mcs}.

The hypercubic crossing equation can be derived by exchanging
$(1,i)\leftrightarrow(3,k)$, collecting terms that multiply the same tensor
structure, and symmetrizing/antisymmetrizing in $u, v$. Defining
\eqn{F_{\Delta,\lsp\ell}^{\pm}(u,v)=v^{\Delta_\phi}g_{\Delta,\lsp\ell}(u,v)
\pm u^{\Delta_\phi}g_{\Delta,\lsp\ell}(v,u)\,,}[Fpmdef]
we find
\eqn{\sum_{\text{S}_\square^+}\lambda_\cO^2\begin{pmatrix}
  0\\
  F^-_{\Delta,\lsp\ell}\\
  F^+_{\Delta,\lsp\ell}\\
  F^-_{\Delta,\lsp\ell}
\end{pmatrix}+
\sum_{\text{V}_\square^+}\lambda_\cO^2\begin{pmatrix}
  0\\
  -\tfrac{2}{N}F^-_{\Delta,\lsp\ell}\\
  -\tfrac{2}{N}F^+_{\Delta,\lsp\ell}\\
  (2-\tfrac{2}{N})F^-_{\Delta,\lsp\ell}
\end{pmatrix}+
\sum_{\text{Y}_\square^+}\!\lambda_\cO^2\begin{pmatrix}
  F^-_{\Delta,\lsp\ell}\\
  F^-_{\Delta,\lsp\ell}\\
  -F^+_{\Delta,\lsp\ell}\\
  0
\end{pmatrix}+
\sum_{\text{A}_\square^{\lnsp-}}\lambda_\cO^2\begin{pmatrix}
  F^-_{\Delta,\lsp\ell}\\
  -F^-_{\Delta,\lsp\ell}\\
  F^+_{\Delta,\lsp\ell}\\
  0
\end{pmatrix}=\begin{pmatrix}
  0\\
  0\\
  0\\
  0
\end{pmatrix}.}[crEqCubic]
Compared to the $O(N)$ case we have one more crossing equation for the
hypercubic theory.

Another way to derive the hypercubic crossing equation is to recall the
presence of a rank-four traceless-symmetric primitive invariant tensor
$d_{ijkl}$ in theories with hypercubic symmetry,
satisfying~\cite{Osborn:2017ucf}
\eqn{d_{ijmn}d_{mnkl}=\frac{N}{(N+2)^2}\Big(
\delta_{ik}\delta_{jl}+\delta_{il}\delta_{jk}-\frac{2}{N}\delta_{ij}
\delta_{kl}\Big) + \frac{N-2}{N+2}\lsp d_{ijkl}\,.}[]
We can then define the linearly-independent invariant projectors
\eqna{P_{ijkl}^{(1)}&=\frac{1}{N}\delta_{ij}\delta_{kl}\,,\\
P_{ijkl}^{(2)}&=d_{ijkl}+\frac{1}{N+2}\Big(
\delta_{ik}\delta_{jl}+\delta_{il}\delta_{jk}-\frac{2}{N}\delta_{ij}
\delta_{kl}\Big)\,,\\
P_{ijkl}^{(3)}&=-d_{ijkl}+\frac{N}{2(N+2)}\Big(
\delta_{ik}\delta_{jl}+\delta_{il}\delta_{jk}-\frac{2}{N}\delta_{ij}
\delta_{kl}\Big)\,,\\
P_{ijkl}^{(4)}&= -\tfrac12(\delta_{ik}\delta_{jl}-
\delta_{il}\delta_{jk})\,,
}[projProp]
that satisfy
\eqn{P_{ijmn}^{(I)}P_{nmkl}^{(J)}=P_{ijkl}^{(I)}\lsp\delta^{IJ}\,,\qquad
\sum_{I}P^{(I)}_{ijkl}=\delta_{il}\delta_{jk}\,,\qquad
P_{ijkl}^{(I)}\lsp\delta_{il}\delta_{jk}=d_r^{(I)}\,,}[]
where $d_r^{(I)}$ is the dimension of the representation indexed by $I$.
The four-point function of $\phi$ can be decomposed in the basis of tensors
$P^{(I)}$, and it is easy to check that the crossing equation that follows
is equivalent to \crEqCubic. Note that the three rank-four projectors of
$O(N)$ are given by $P^{(1)}_{ijkl}, P^{(2)}_{ijkl}+P^{(3)}_{ijkl}$, and
$P^{(4)}_{ijkl}$.

\newsec{Bounds in hypercubic theories}[boundsHyperC]
We are now ready to obtain bounds using \crEqCubic. In this work we use
\texttt{PyCFTBoot}~\cite{Behan:2016dtz} to produce the input for
\texttt{SDPB}~\cite{Simmons-Duffin:2015qma}, which performs the numerical
optimization. Unless otherwise noted, for the plots of this paper we use
$\texttt{nmax}=9$, $\texttt{mmax}=6$, $\texttt{kmax}=36$,
$\texttt{cutoff}=10^{-10}$ in \texttt{PyCFTBoot}, and we include spins up
to $\ell_{\text{max}}=26$. For \texttt{SDPB} we use the options
\texttt{--findPrimalFeasible} and \texttt{--findDualFeasible},\foot{With
these options if \texttt{SDPB} finds a primal feasible solution then the
assumed operator spectrum is allowed, while if it finds a dual feasible
solution then the assumed operator spectrum is excluded.} and we set
$\texttt{precision}=660$, $\texttt{dualErrorThreshold}=10^{-20}$ and
default values for other parameters. We have found that with these choices
we obtain $O(N)$ bounds that look identical to those of~\cite{Kos:2013tga}.
The bounds are obtained with a vertical tolerance of $10^{-3}$.

\begin{figure}[ht]
  \centering
  \includegraphics{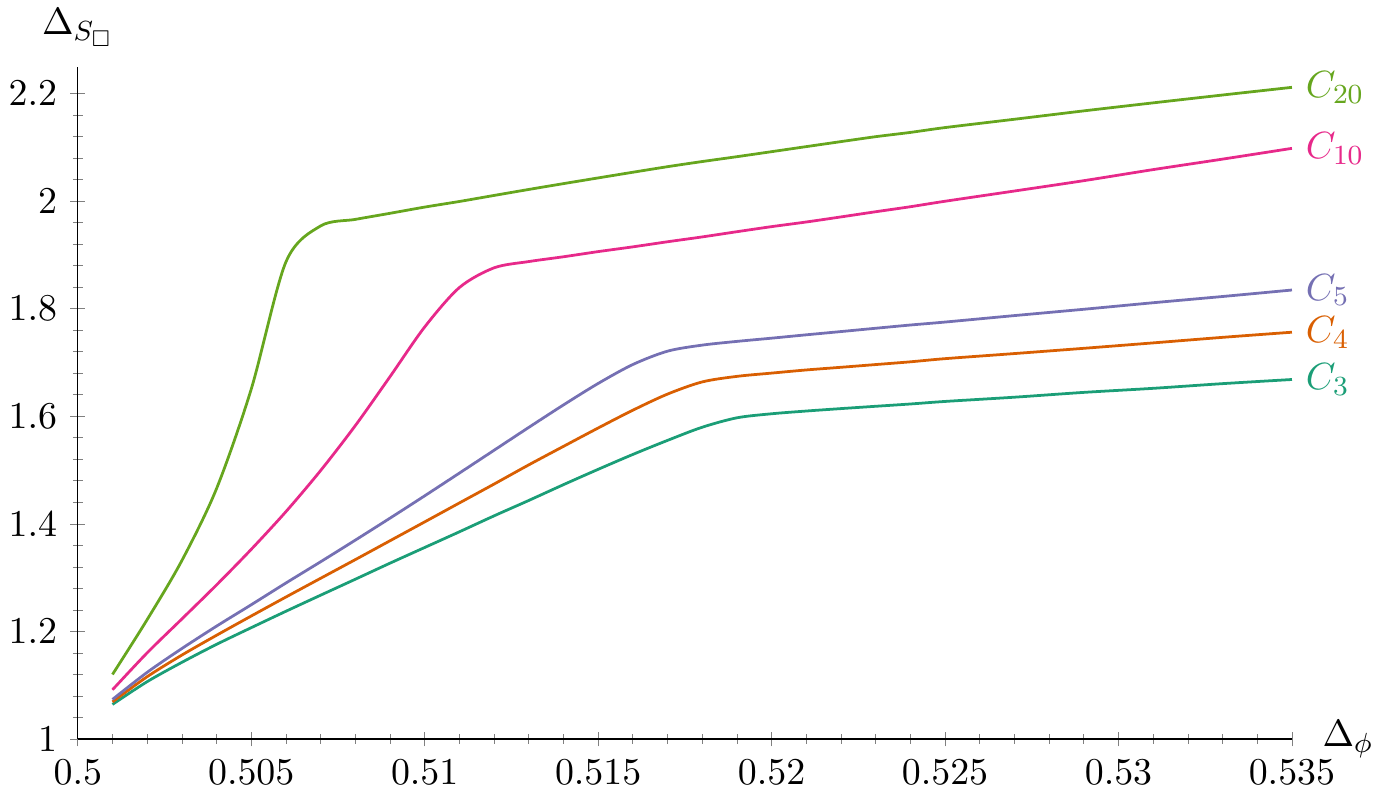}
  \caption{Upper bound on the dimension of the first singlet scalar in the
  $\phi_i\times\phi_j$ OPE as a function of the dimension of $\phi$. Areas
  above the curves are excluded in the corresponding theories.}
  \label{fig:Delta_S_cubic}
\end{figure}

\subsec{Operator dimension bounds}
A bound on the dimension of the first singlet scalar in the OPE of
$\phi_i\times\phi_j$ is shown in Fig.~\ref{fig:Delta_S_cubic}.
The high degree of similarity of Fig.~\ref{fig:Delta_S_cubic} with the
$O(N)$ bounds of~\cite{Kos:2013tga} suggests that although we only require
hypercubic symmetry, the bound on $\Delta_{S_\square}$ is saturated by the
$O(N)$ solution. For $N=3$ we have explicitly checked that the bound
obtained from the crossing equation \crEqCubic is exactly the same as that
obtained from the $O(3)$ crossing equation. In the singlet sector,
therefore, the $O(N)$ solution gets in the way and does not allow
saturation of the bound with purely hypercubic theories, which lie
somewhere in the allowed region.

We now move on to the $\text{V}_\square^+$ sector. A bound on the first
scalar operator in that sector, called $V_\square$ here, is shown in
Fig.~\ref{fig:Delta_V_cubic}.\foot{From a weakly-coupled point of view the
operator $V_\square$ is of the form
$(\delta_{ijkl}-\frac{1}{N}\delta_{ij}\delta_{kl})\phi_k\phi_l$~\cite{Dey:2016mcs}.}
\begin{figure}[ht]
  \centering
  \includegraphics{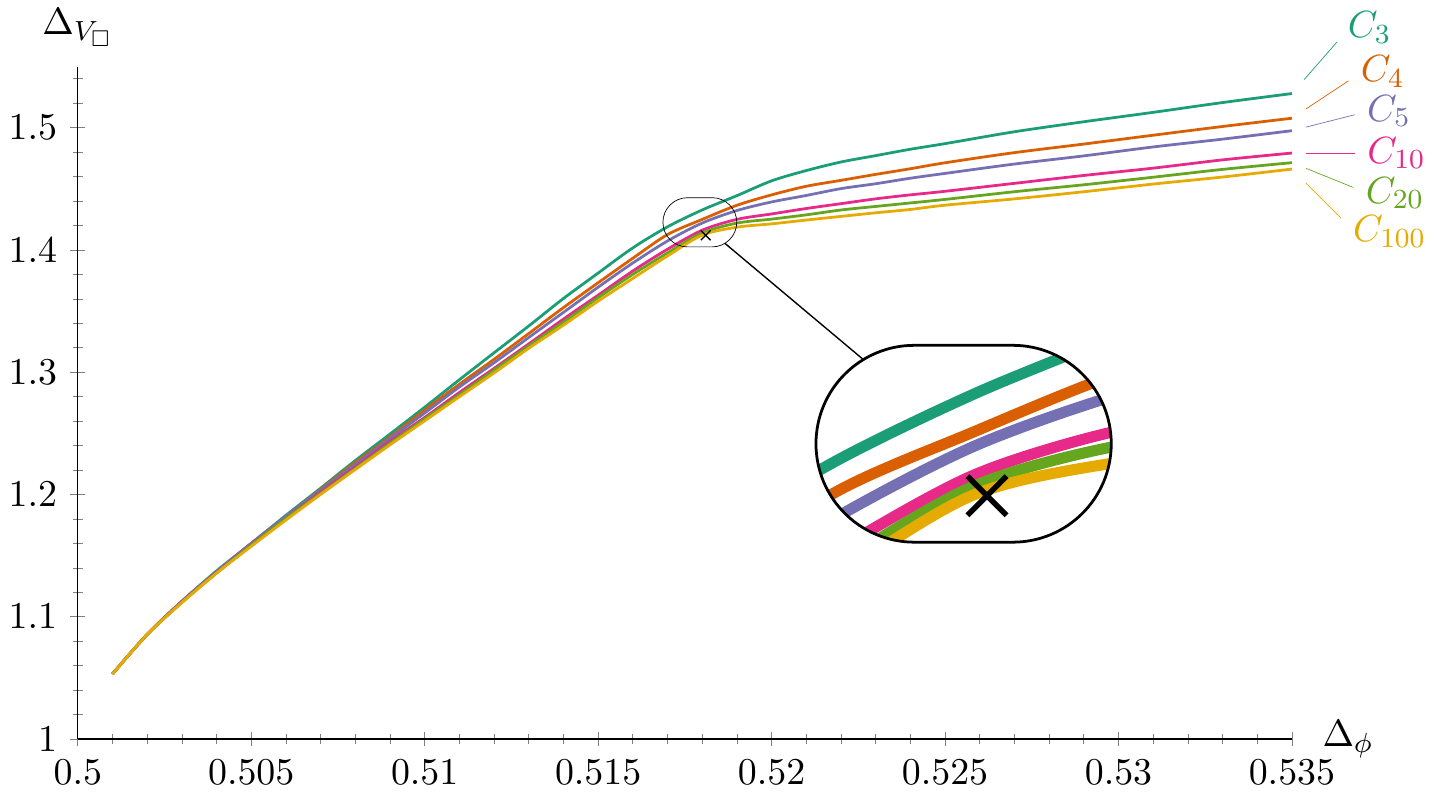}
  \caption{Upper bound on the dimension of the first scalar operator in the
  $\text{V}^+_\square$ sector of the $\phi_i\times\phi_j$ OPE as a function
  of the dimension of $\phi$.  Areas above the curves are excluded in the
  corresponding theories. The x-marker indicates the position of the
  decoupled Ising theory.}
  \label{fig:Delta_V_cubic}
\end{figure}
Perturbation theory gives us two hypercubic fixed points: one is fully
interacting, while the other is obtained by taking $N$ decoupled Ising
models~\cite{Osborn:2017ucf}.  In the latter case $V_\square$ has the
scaling dimension of the $\epsilon$ operator in the Ising model,
$\Delta_\epsilon\approx1.4126$. While the putative theory that saturates
the bound of Fig.~\ref{fig:Delta_V_cubic} for small $N$ is unknown to us at
this point, we can easily see that it is not the decoupled Ising one.  Of
course the latter is always in the allowed region, and in fact at large $N$
the bound gets closer to it.

\begin{figure}[H]
  \centering
  \includegraphics{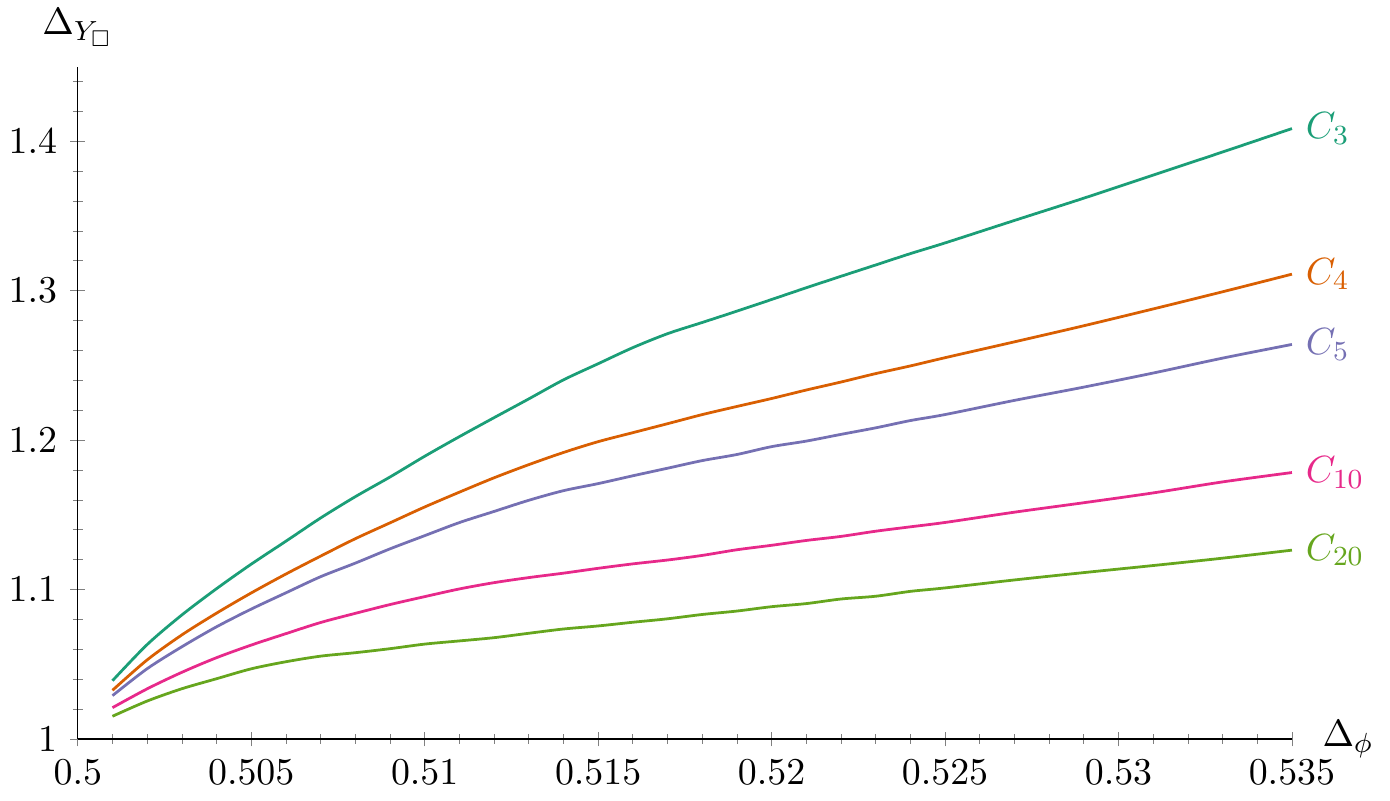}
  \caption{Upper bound on the dimension of the first scalar operator in the
  $\text{Y}^+_\square$ sector of the $\phi_i\times\phi_j$ OPE as a function
  of the dimension of $\phi$.  Areas above the curves are excluded in the
  corresponding theories.}
  \label{fig:Delta_Y_cubic}
\end{figure}

\begin{figure}[H]
  \centering
  \includegraphics{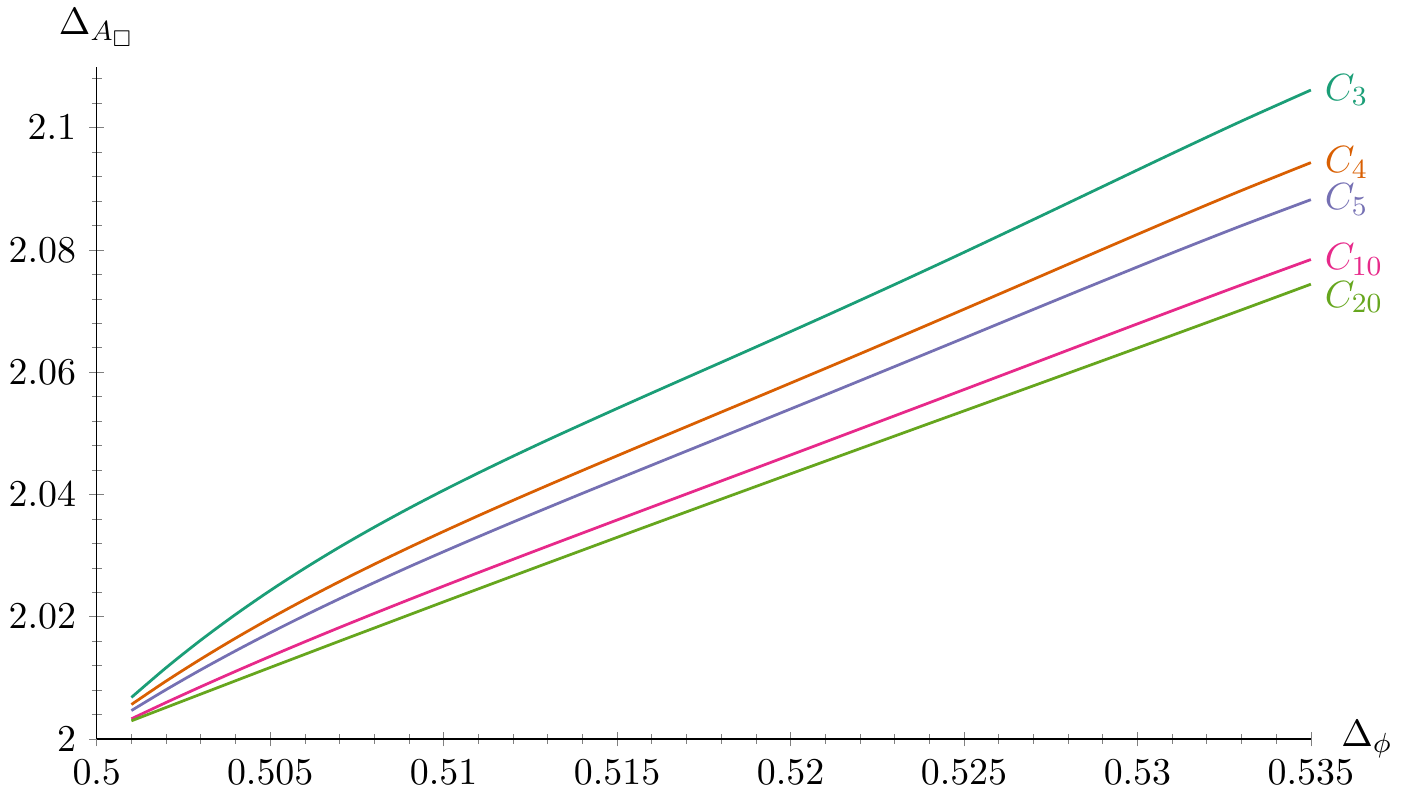}
  \caption{Upper bound on the dimension of the first vector operator in the
  $\text{A}^-_\square$ sector of the $\phi_i\times\phi_j$ OPE as a function
  of the dimension of $\phi$.  Areas above the curves are excluded in the
  corresponding theories.}
  \label{fig:Delta_A_cubic}
\end{figure}

We can also obtain bounds for the first scalar operator in the
$\text{Y}_\square^+$ sector and the first vector operator in the
$\text{A}_\square^-$ sector. These bounds are shown in
Figs.~\ref{fig:Delta_Y_cubic} and \ref{fig:Delta_A_cubic}, respectively.
Unfortunately, they do not show any interesting features. In
Fig.~\ref{fig:Delta_A_cubic} the genaralized free theory line
$2\Delta_\phi+1$ is of course in the allowed region of all bounds.  Note
that because the hypercubic symmetry is discrete we do not have a conserved
current in the $\text{A}_\square^-$ sector.

To recover the $O(N)$ solution the $V_\square$ and $Y_\square$ operators
should have the same scaling dimension and combine to the lowest-dimension
scalar traceless-symmetric operator of $O(N)$ with dimension
$\Delta_{T_{\mcirc}}$. That possibility should be allowed by our bounds in
Figs.~\ref{fig:Delta_V_cubic} and~\ref{fig:Delta_Y_cubic}, i.e.\
$\Delta_{T_{\mcirc}}$ should be in the allowed region of the corresponding
bound in both Figs.~\ref{fig:Delta_V_cubic} and~\ref{fig:Delta_Y_cubic}. It
is immediately verified based on the results of~\cite{Kos:2013tga,
Kos:2015mba} that this is indeed the case.

Some of our bounds can be compared with the $\veps$ expansion using the
three-loop result, see e.g.~\cite{Kleinert:1994td} or~\cite{Dey:2016mcs},
\eqn{\Delta_\phi^\veps=1-\tfrac12\lsp\veps+\frac{(N-1)(N+2)}{108N^2}\lsp\veps^2
+\frac{(N-1)(109N^3-222N^2+1728N-1696)}{11664N^4}\lsp\veps^3+\text{O}(\veps^4)\,,}[]
and the two-loop results of~\cite{Dey:2016mcs},
\eqna{\Delta_{V_\square}^\veps&=2-\frac{2(N+1)}{3N}\lsp\veps
+\frac{19N^3+131N^2-538N+424}{162N^3}\lsp\veps^2+\text{O}(\veps^3)\,,\\
\Delta_{Y_\square}^\veps&=2-\frac{3N-2}{3N}\lsp\veps
+\frac{3N^3-127N^2+530N-424}{162N^3}\lsp\veps^2+\text{O}(\veps^3)\,.}[]
As an example, choosing $N=3$ and sending $\veps\to1$ neglecting
$\text{O}(\veps^4)$ or $\text{O}(\veps^3)$ terms we find
$\Delta_\phi^{\veps=1}\approx 0.52$, $\Delta_{V_\square}^{\veps=1}\approx
1.23$ and $\Delta_{Y_\square}^{\veps=1}\approx 1.25$. These values are in
the allowed regions of the corresponding bounds in
Fig.~\ref{fig:Delta_V_cubic} and Fig.~\ref{fig:Delta_Y_cubic}. At large $N$
$\Delta_\phi^{\veps=1}\approx 0.52$, $\Delta_{V_\square}^{\veps=1}\approx
1.45$, and $\Delta_{Y_\square}^{\veps=1}\approx 1.02$, so in that case
$\Delta_{V_\square}^{\veps=1}$ is in the excluded region of
Fig.~\ref{fig:Delta_V_cubic}. Of course these $\veps=1$ results should be
taken with a grain of salt. Higher-order results for
$\Delta_{V_\square}^\veps$ and $\Delta_{Y_\square}^\veps$ would allow
proper resummations and the extraction of more meaningful results.

\subsec{Analysis of the spectrum of the \texorpdfstring{$C_3$}{C3} boundary
solution}
So far we have applied the standard numerical bootstrap logic for operator
dimension bounds. Let us summarize it briefly and schematically here.
First, we bring the crossing equation to the form
\eqn{\sum_{\text{all sectors}}\!\!\!\!\lambda_{\cO}^2\lsp
\vec{V}_{\Delta,\lsp\ell} =-\vec{V}_{0,0}\,,}[schemCrEq]
where in the right-hand side we have isolated the contribution of the
identity operator, whose OPE coefficient has been normalized to unity. The
vectors $\vec{V}$ stand for the various contributions in \crEqCubic for
example.  Now we take a linear functional $\vec{\alpha}$ of appropriate
dimension and form the inner product with \schemCrEq, i.e.\ we write
\eqn{\sum_{\text{all sectors}}\!\!\!\!\lambda_{\cO}^2\,
\vec{\alpha}\cdot \vec{V}_{\Delta,\lsp\ell}
=-\vec{\alpha}\cdot\vec{V}_{0,0}\,.}[funcAction]
At this point we make an assumption on the spectrum and we scan over the
space of linear functionals $\vec{\alpha}$ demanding
\eqn{\vec{\alpha}\cdot\vec{V}_{0,0}=1\,,\qquad
\vec{\alpha}\cdot\vec{V}_{\Delta,\lsp\ell}\ge0\,,\text{ for all
allowed }\Delta,\lsp\ell.}[assums]
If we manage to find such a functional, then \funcAction leads to a
contradiction and so the assumption we made on the spectrum is not
consistent with unitarity (i.e.\ with assuming that all $\lambda_{\cO}$'s
are real).

Right at the boundary of the allowed region knowledge of the functional is
enough to give us information about the spectrum of the actual solution to
crossing symmetry there~\cite{Poland:2010wg, ElShowk:2012hu}. This is
because when we cross the bound we go from having a functional to not
having one, which means that right on the boundary of the allowed and the
disallowed region (on the disallowed side) the action of the associated
extremal functional should saturate the inequalities in \assums and give
zero on $\vec{V}_{\Delta,\lsp\ell}$ for all physical operators in the
spectrum.  In other words, right on the bound (on the disallowed side) the
left-hand side of \funcAction is zero because all contributions (at
discrete $\Delta$'s for allowed $\ell$'s) are zero.\foot{Of course some OPE
coefficients could also become zero, but we do not expect this to happen
away from the unitarity bounds.} Note that the right-hand side of
\funcAction is equal to $-1$ due to \assums. The functional we obtain is
thus truly extremal when for generic $\Delta,\ell$ outside the spectrum
$\vec{\alpha}\cdot\vec{V}_{\Delta,\lsp\ell}/\vec{\alpha}\cdot\vec{V}_{0,0}\to\infty$.

Our results below are extracted by plotting the logarithm of the action of
the functional on the convolved conformal blocks\foot{To obtain these plots
we made some minor additions to \texttt{PyCFTBoot} and used
\texttt{Matplotlib}~\cite{Hunter:2007}.} and looking at the positions of
the dips in those plots.\foot{To find the positions of the dips we used
\texttt{WebPlotDigitizer}~\cite{WebPlotDigitizer}.} These give us the
scaling dimensions of operators that solve crossing at the bound.

Let us first obtain the functional along the bound of
Fig.~\ref{fig:Delta_S_cubic}. From the spectrum in the
$\text{V}_\square^+$, $\text{Y}_{\square}^+$ and $\text{A}_\square^-$
sectors we can verify that our $\Delta_{S_\square}$ bound is saturated by
the $O(N)$ solution. Indeed, operators in the $\text{V}_\square^+$ and
$\text{Y}_\square^+$ sectors have the same scaling dimension, and the first
operator in the $\text{A}_\square^-$ sector has dimension exactly equal to
two.

We will now perform an analysis of the spectrum along the $C_3$ bound using
the functional obtained from Fig.~\ref{fig:Delta_V_cubic}. We will refer to
the spectrum at the kink as the $C_{3}^b$ solution. Note that we have
obtained the bound of Fig.~\ref{fig:Delta_V_cubic} with a vertical
tolerance of $10^{-6}$ for the spectrum analysis that follows. We would
like to mention that the results reported in the remainder of this section
do not change by making the numerics more demanding. We have checked this
by running with $\texttt{nmax}=11$, $\texttt{mmax}=8$, $\texttt{kmax}=36$,
$\texttt{cutoff}=10^{-12}$ in \texttt{PyCFTBoot}, and including spins up to
$\ell_{\text{max}}=32$.

\begin{figure}[ht]
  \centering
  \includegraphics{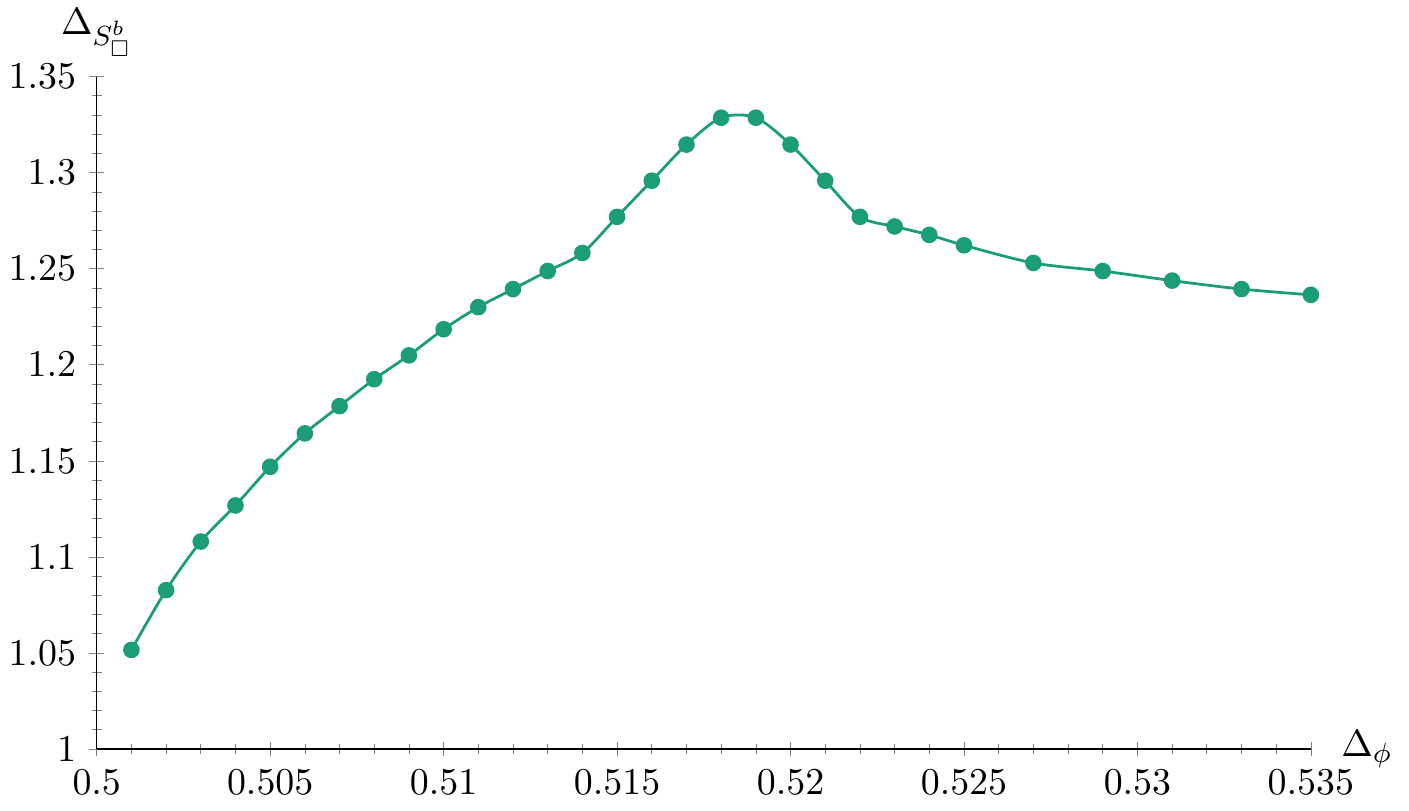}
  \caption{The dimension of $S_\square^b$ at the $C_3^b$ solution, i.e.\
  assuming that $\Delta_{V_\square}$ is equal to the bound of
  Fig.~\ref{fig:Delta_V_cubic}. The markers indicate the points at which we
  have computed the spectrum.} \label{fig:Delta_S_from_spectrum_cubic}
\end{figure}

First, we look at the dimension of $S_\square^b$ in the $C_3^b$ solution in
Fig.~\ref{fig:Delta_S_from_spectrum_cubic}. Our result is that the $C_3^b$
solution at the kink, occurring at $\Delta_\phi\approx 0.518$, has much
lower $\Delta_{S_\square^b}$ than $\Delta_{S_{\mcirc}}$ at the $O(3)$ fixed
point.  Indeed, we find $\Delta_{S_\square^b}\approx 1.329$,
while~\cite{Kos:2016ysd} gives $\Delta_{S_{\mcirc}}\approx 1.5957$.  This
suggests that the $C_3^b$ solution does not correspond to the $C_3^\veps$
theory found with the $\veps$ expansion, where the scaling dimensions
$\Delta_{S_\square^\veps}$ and $\Delta_{S_{\mcirc}}$ at $C_3^\veps$ and
$O(3)$, respectively, are calculated to be nearly
degenerate~\cite{Aharony:1973zz, Carmona:1999rm, Pelissetto:2000ek}.
We would also like to note here that we now have a further indication that
the bound in Fig.~\ref{fig:Delta_V_cubic} is not saturated by the decoupled
Ising theory. Indeed, if that were the case then the dimension of
$S_\square^b$ should be equal to that of the $\epsilon$ operator in the
Ising model, which is clearly not the case based on
Fig.~\ref{fig:Delta_S_from_spectrum_cubic}.

\begin{figure}[H]
  \centering
  \includegraphics{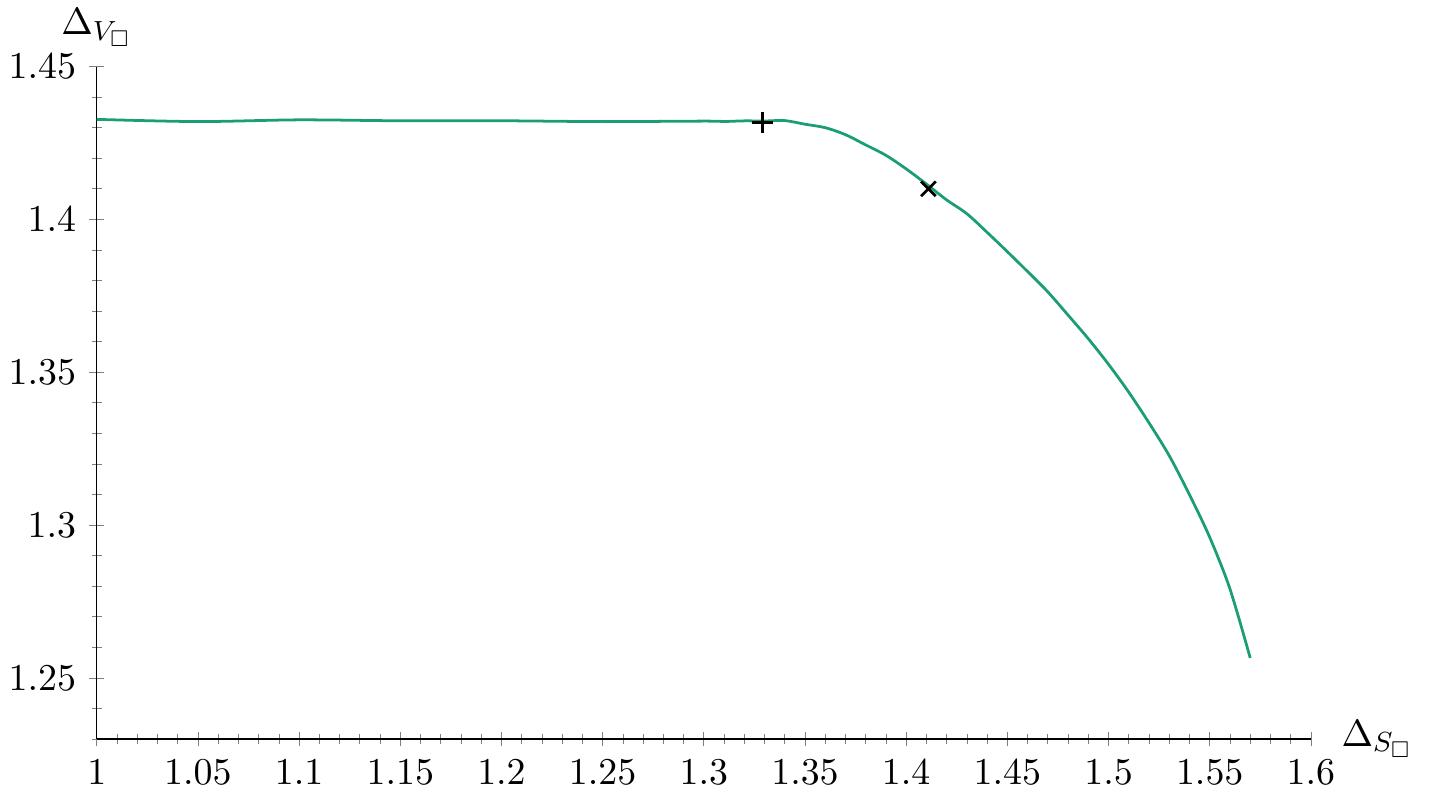}
  \caption{Upper bound on $\Delta_{V_\square}$ as a function of the gap on
  $\Delta_{S_\square}$ assuming $\Delta_\phi=0.518$ for $N=3$. The x- and
  +-markers indicate the position of the decoupled Ising and the $C_3^b$
  solution, respectively.} \label{fig:Delta_S-Delta_V_bound_cubic}
\end{figure}

We can also fix $\Delta_\phi$ to its value at the kink,
$\Delta_\phi=0.518$, and obtain a bound on $\Delta_{V_\square}$ by imposing
a gap on $\Delta_{S_\square}$ in the allowed region of
Fig.~\ref{fig:Delta_S_cubic}. That is, instead of allowing
$\Delta_{S_\square}$ above the unitarity bound, we allow it only above the
value in the figure.  This bound is shown in
Fig.~\ref{fig:Delta_S-Delta_V_bound_cubic}.

\begin{figure}[H]
  \centering
  \includegraphics{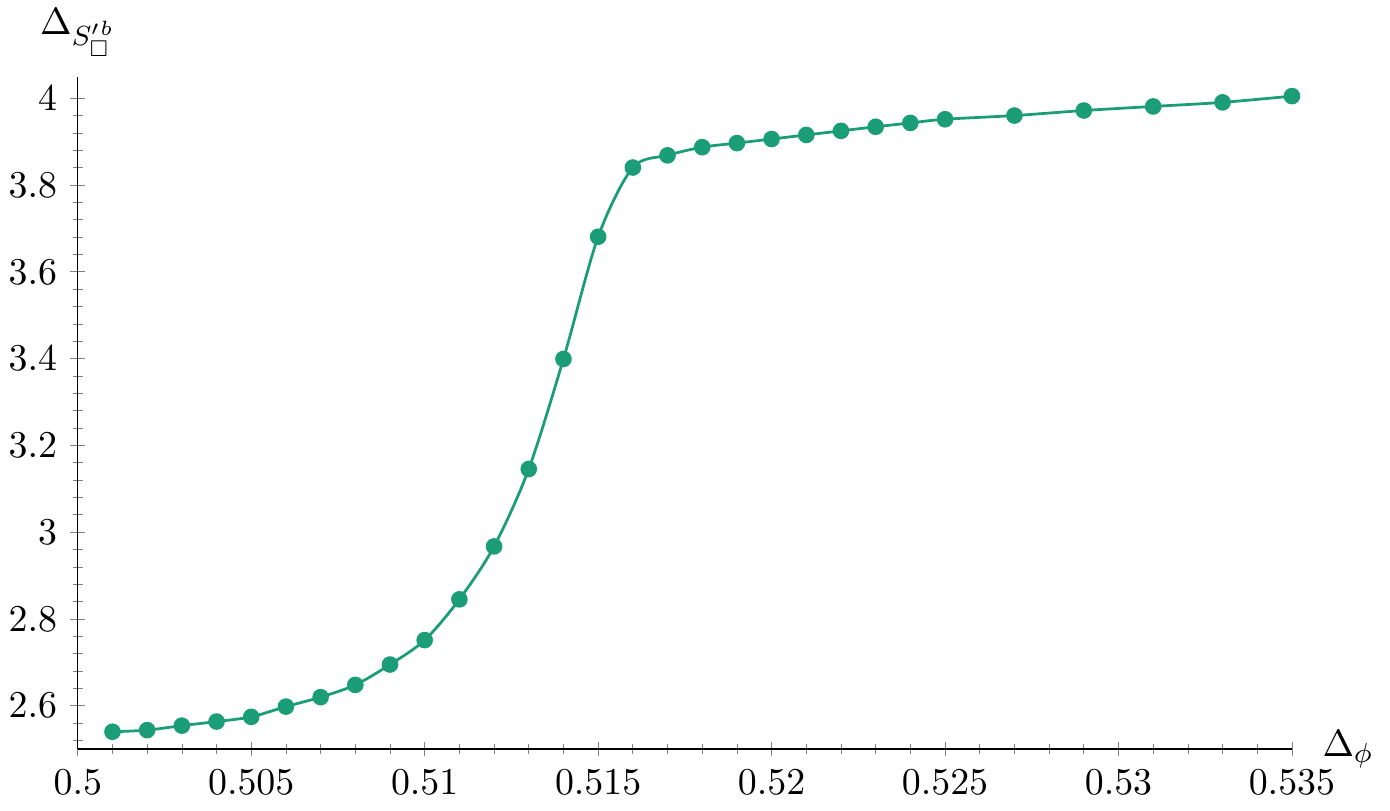}
  \caption{The dimension of $S_\square^{\prime\llsp b}$ at the
    $C_3^b$ solution, i.e.\ assuming that $\Delta_{V_\square}$ is equal to
    the bound of Fig.~\ref{fig:Delta_V_cubic}. The markers indicate the
    points at which we have computed the spectrum.}
    \label{fig:Delta_Sp_from_spectrum_cubic}
\end{figure}

We may also obtain the dimension of the second scalar singlet in the
$\phi_i\times\phi_j$ OPE, called $S_\square^{\prime\llsp b}$ here. If the
scaling dimension of this operator were less than three, then $C_3^b$ would
be a tricritical solution (two relevant scalar singlets).  From
Fig.~\ref{fig:Delta_Sp_from_spectrum_cubic} we see that
$\Delta_{S_\square^{\prime\llsp b}}>3$ around $\Delta_\phi=0.518$, and so
$C_3^b$ is a critical solution (one relevant scalar singlet).

\begin{figure}[H]
  \centering
  \includegraphics{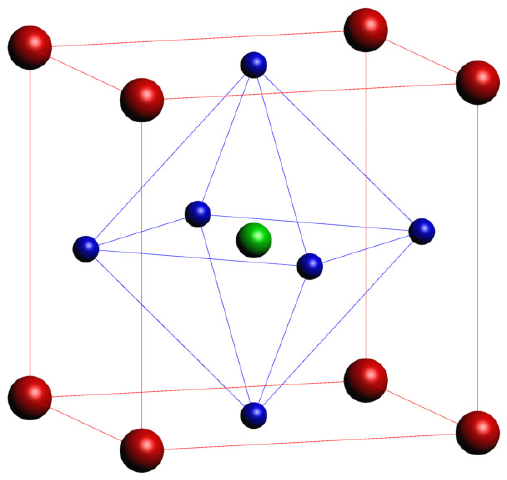}
  \caption{The perovskite structure ABX$_3$. A is red, B is green, and
  X$_3$ is blue.} \label{fig:perovskite}
\end{figure}

\subsec{Discussion}
Our results may have implications for the theory of structural phase
transitions. It was suggested a long time ago~\cite{Aharony:1973zz} that
critical exponents of the $C_3^\veps$ theory should be compared to
experiment, for example the structural phase transition of $\text{SrTiO}_3$
(strontium titanate)~\cite{PhysRevLett.26.13, RISTE19711455,
PhysRevLett.28.503, PhysRevB.7.1052, CowShap} where the lattice structure
undergoes a transition from cubic to tetragonal at a critical temperature
$T_c\approx 100\text{ K}$. Crystals of the type ABX$_3$ are called
perovskites. X is usually oxygen. The undistorted (high-temperature) phase
has A on the corners of a cube, B at the base center, and X$_3$ at the face
centers; see Fig.~\ref{fig:perovskite}.

\begin{figure}[ht]
  \centering
  \includegraphics{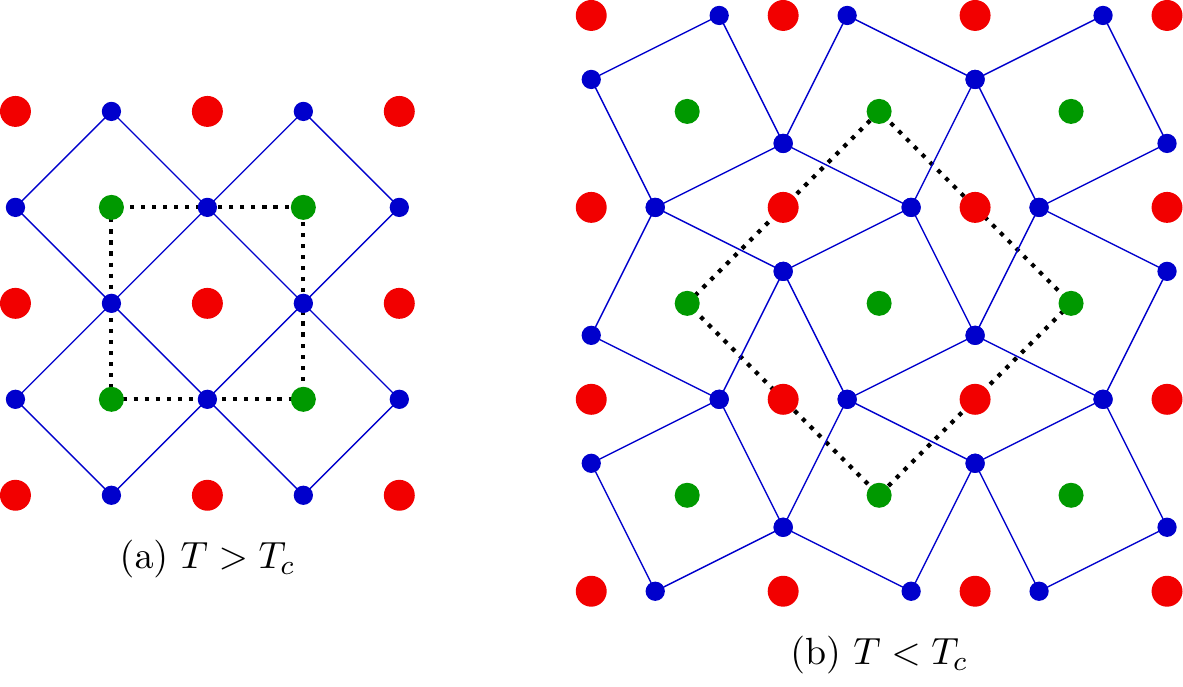}
  \caption{The crystal structure in a top-down view above (a) and below (b)
  the critical transition temperature $T_c\approx 100\text{ K}$. The unit
  cell is highlighted by the dotted line.} \label{fig:cubic_to_tetragonal}
\end{figure}

In these systems, and in the context of the Landau theory of phase
transitions, a term preserving cubic symmetry is allowed in the expansion
of the free energy, and so its effects need to be taken into
consideration~\cite[sec.\ I.4.2]{Cowley}. A transition to a distorted phase
occurs due to rotations of the X$_3$ octahedra around a four-fold axis of
the cube as the temperature is lowered below a critical value. The phase
transition is continuous (second order), in the sense that the rotation of
the octahedra is continuous. The crystal structure above and below the
transition temperature is depicted in Fig.~\ref{fig:cubic_to_tetragonal} in
a top-down view.
In the third direction two adjacent octahedra rotate in opposite directions
and so the unit cell is enlarged by a factor of 2 in that direction. Taking
everything into account the unit cell of the distorted phase is enlarged by
$\sqrt{2}\times\sqrt{2}\times 2$ relative to the undistorted phase, and so
it belongs to the tetragonal crystal system. The symmetry of the
undistorted phase is given by the group $O_h$ (which is the same as what we
call $C_3$), while that of the distorted phase is given by the 16-element
group $D_{4h}$. A review on structural phase transitions is~\cite{Cowley,
Bruce}, while some information can also be found in~\cite[Chapter
XIV]{Landau:1980mil}.

As we already mentioned, the $\veps$ expansion gives
$\Delta_{S_\square^\veps}$ very close to
$\Delta_{S_{\mcirc}}$~\cite{Aharony:1973zz, Carmona:1999rm,
Pelissetto:2000ek}, which was already noticed in~\cite{Aharony:1973zz} as a
possible disagreement with experiment.  It was later suggested that
residual strains in the crystals used in the experiments may be responsible
for a crossover to Ising-like behavior~\cite{PhysRevLett.33.427}.\foot{The
presence of strains brings about new terms in the expansion of the free
energy in the context of the Landau theory of phase transitions.} The Ising
critical exponents match the experimental results very well, thus offering
a way out of the apparent incompatibility between the $\veps$ expansion
results and the experiments.  The suggestion that systematic strains induce
an Ising-like behavior was investigated in subsequent
experiments~\cite{PhysRevLett.35.1547}. To our knowledge critical exponents
in strain-free crystals have not been reported, although such results are
mentioned in~\cite{CowShap} (presumably pertaining to
\cite{PhysRevB.6.4332}) to be close to those obtained in crystals with
strains. In light of our results it would be very interesting to revisit
these experiments in order to see if our $C_3^b$ solution corresponds to
the theory describing structural phase transitions of strain-free
perovskites.

Our results also imply that the $C_3^b$ solution is not relevant for cubic
magnets, whose critical exponents have been measured and found close to
those predicted by the $O(3)$ model and the $C_3^\veps$
theory~\cite{Pelissetto:2000ek}. This is despite the fact that the $C_3^b$
solution has only one relevant scalar singlet, as we deduce from
Fig.~\ref{fig:Delta_Sp_from_spectrum_cubic}. We do not know if the cubic
deformation is relevant in the $O(3)$ theory. Note that in the $O(N)$ model
the cubic deformation is drawn from a traceless-symmetric scalar operator
with four $O(N)$ indices, $\cO_{ijkl}$~\cite{Pelissetto:2000ek}. In the
$\veps$ expansion it is known that, when it exists, the stable fixed point
is unique~\cite{Michel:1983in}.\foot{We reiterate that by ``stable'' we
mean that the only relevant deformation is the mass.}  If the $C_3^b$
solution were to correspond to the $C_3^\veps$ theory, then the
non-perturbative situation would contradict that intuition. On the one hand
cubic magnets would be described by the $O(3)$ theory at criticality
according to the experimental results.  This would mean that at the $O(3)$
fixed point the cubic deformation is not relevant.  On the other hand the
bootstrap would be showing that $C_3^b$ has just one relevant operator and
so it would correspond to a stable fixed point. Despite that, cubic magnets
at criticality would not be in this universality class.

We would then be forced to conclude that neither the $C_3^b$ or $O(3)$
fixed point has a second relevant operator (assuming that the critical
exponents of cubic magnets have been measured correctly). Besides the
possibility that both fixed points are stable, it could also be that the
$O(3)$ model has an exactly marginal operator.  A Monte Carlo study of the
cubic deformation at the $O(3)$ fixed point was performed
in~\cite{Caselle:1997gf}. Their results for the anomalous dimension of the
cubic deformation are consistent with the value zero.  If zero were indeed
the correct answer, this would obviously imply that the cubic deformation
is exactly marginal at the $O(3)$ fixed point.  The necessary condition is
that a traceless-symmetric scalar operator $\cO_{ijkl}$ of the $O(3)$
theory has dimension exactly equal to three. To our knowledge this has not
been excluded in the literature, but without a principle that would fix the
dimension of $\cO_{ijkl}$ to three it seems unlikely.

\subsec{Central charge bounds}
Without any assumptions we can bound the central charge $C_\square$ as a
function of $\Delta_\phi$. Here we present lower bounds on the ratio
$C_\square/C_{\text{free}}$, where $C_{\text{free}}=\frac32 N$. We remind
the reader that $C$ appears in the coefficient of the two-point function of
the stress-energy tensor, which in $d$ dimensions is constrained by
conformal invariance to be of the form
\eqn{\langle T_{\mu\nu}(x)\llsp T_{\rho\sigma}(0)\rangle=
  C\frac{1}{S_d^{\llsp 2}}\frac{1}{(x^2)^d}\lsp
\mathcal{I}_{\mu\nu\rho\sigma}(x)\,,}[]
where $S_d=2\pi^{\frac12 d}/\Gamma(\frac12 d)$ and
\eqn{\qquad\mathcal{I}_{\mu\nu\rho\sigma}=\tfrac12(I_{\mu\rho}\lsp
I_{\nu\sigma}+I_{\mu\sigma}\lsp
I_{\nu\rho})-\frac{1}{d}\lsp\eta_{\mu\nu}\eta_{\rho\sigma}\,,
\qquad I_{\mu\nu}=\eta_{\mu\nu}-\frac{2}{x^2}\lsp x_\mu x_\nu\,.}[]
In these conventions a free scalar's contribution to the central charge is
equal to $d/(d-1)$.

\begin{figure}[H]
  \centering
  \includegraphics{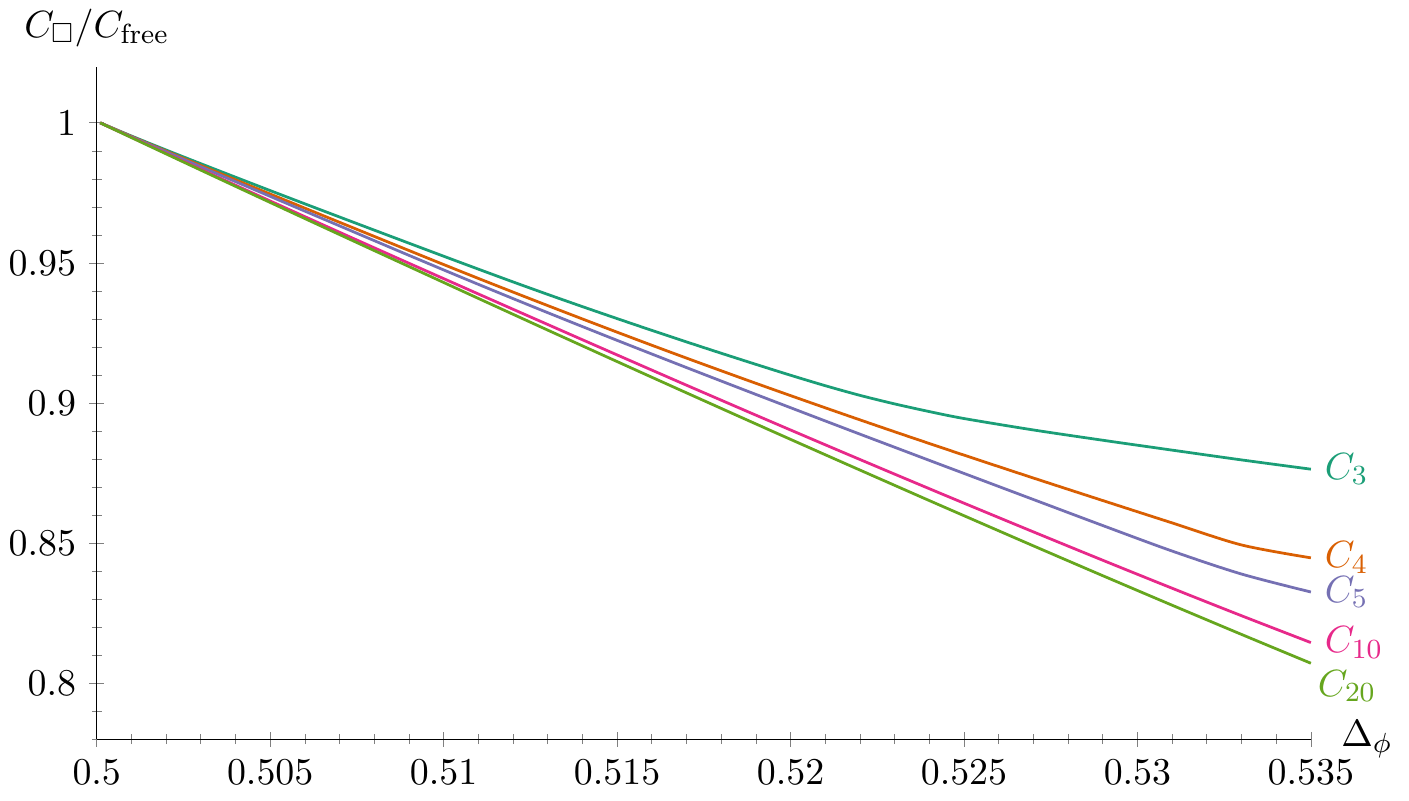}
  \caption{Lower bound on the central charge as a function of the dimension
  of $\phi$. Areas below the curves are excluded in the corresponding
  theories.}
  \label{fig:cc}
\end{figure}

Without any assumptions the bounds are shown in Fig.~\ref{fig:cc}. They are
essentially identical to the ones obtained for the $O(N)$
models~\cite{Kos:2013tga}.

\begin{figure}[H]
  \centering
  \includegraphics{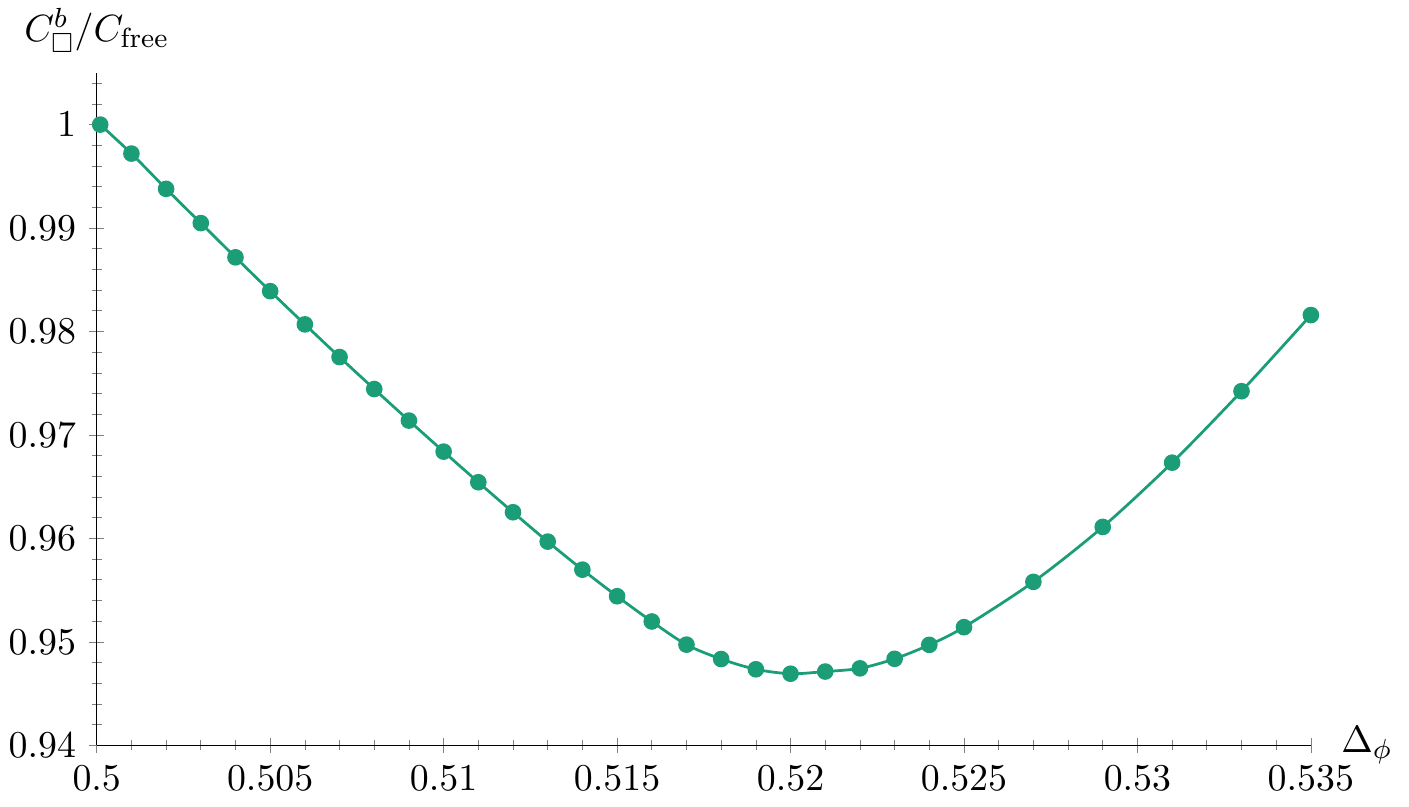}
  \caption{Value of the central charge of the $C_3^b$ solution as a
  function of the dimension of $\phi$.}
  \label{fig:cc_with_V_at_bound_cubic}
\end{figure}

For $C_3$ we may assume, as we have before, that $\Delta_{V_\square}$
saturates the bound of Fig.~\ref{fig:Delta_V_cubic} (obtained with a
vertical tolerance of $10^{-6}$). The value of $C_\square^b$ is then shown
in Fig.~\ref{fig:cc_with_V_at_bound_cubic}. Much like in the $O(N)$ and
Ising models there is a minimum. As we see it is rather wide and is
attained slightly to the left of $\Delta_\phi=0.52$. Based on
Fig.~\ref{fig:cc_with_V_at_bound_cubic} we may conclude that the $C_3^b$
solution has a central charge slightly higher than the $O(3)$
value~\cite{Kos:2013tga}, namely $C_{\square}^b\approx 0.947\times
3\times\frac32$, while $C_{\mcirc}\approx 0.944\times 3\times\frac32$.

\newsec{Hypertetrahedral symmetry}[hyperTetra]
Hypertetrahedral symmetry may be implemented by considering $N+1$ vectors
$e_i{\!}^\alpha$, $\alpha =1,2 , \dots , N+1$, forming the vertices of an
$N$-dimensional hypertetrahedron, satisfying
\eqn{{\sum_\alpha} \, e_i{\!}^\alpha  = 0 , \, \qquad
{\sum_\alpha} \, e_i{\!}^\alpha e_j{\!}^\alpha = \delta_{ij} \, , \qquad
e_i{\!}^\alpha e_i{\!}^\beta =\delta^{\alpha\beta} - \frac{1}{N+1}
\equiv P^{\alpha\beta}\,.}[]
In this case the symmetry group is $T_N=\cS_{N+1}$, and we can borrow some
results from the literature. Crucial representation theory was worked out
in~\cite{Wallace:1978hn, Remmel:1989dl}.  In~\cite{Hogervorst:2016itc} the
four-point function of the field $\phi^\alpha=e_i{\!}^\alpha\phi_i$ was
given. Using results of~\cite{Vasseur:2013baa, Couvreur:2017inl} one can
indeed write, in the $12\rightarrow34$ channel,
\eqna{x_{12}^{2\Delta_\phi}x_{34}^{2\Delta_\phi}
\langle\phi^\alpha(x_1)\phi^\beta(x_2)\phi^\gamma(x_3)
\phi^\delta(x_4)\rangle &=
\sum_{\text{S}^+_\tetra} \lambda_\cO^2\lsp
P^{\alpha\beta}P^{\gamma\delta} g_{\Delta,\lsp\ell}(u,v)
+\sum_{\text{V}^+_\tetra} \lambda_\cO^2\lsp
Q^{\alpha\beta\gamma\delta} \lsp g_{\Delta,\lsp\ell}(u,v)\\
&\quad+\sum_{\text{Y}^+_\tetra} \lambda_\cO^2\lsp
R^{\alpha\beta\gamma\delta} g_{\Delta,\lsp\ell}(u,v)
+\sum_{\text{A}^-_\tetra} \lambda_\cO^2\lsp S^{\alpha\beta\gamma\delta}
g_{\Delta,\lsp\ell}(u,v)\,,
}[tetraFour]
where
\eqna{Q^{\alpha\beta\gamma\delta}&=\Big(\delta^{\alpha\beta}
-\frac{2}{N+1}\Big)\Big(\delta^{\gamma\delta}-\frac{2}{N+1}\Big)
\Big(\delta^{\alpha\gamma}+\delta^{\alpha\delta}+\delta^{\beta\gamma}
+\delta^{\beta\delta}-\frac{4}{N+1}\Big)\,,\\
R^{\alpha\beta\gamma\delta}&=\delta^{\alpha\ne\beta}\delta^{\gamma\ne\delta}
\Big(\delta^{\alpha\gamma}\delta^{\beta\delta}+ \delta^{\alpha\delta}
\delta^{\beta\gamma}-\frac{1}{N-1}(\delta^{\alpha\gamma}+
\delta^{\alpha\delta} +\delta^{\beta\gamma}+\delta^{\beta\delta})
+\frac{2}{N(N-1)}\Big)\,,\\
S^{\alpha\beta\gamma\delta}&=\delta^{\alpha\gamma}\delta^{\beta\delta}
-\delta^{\alpha\delta}\delta^{\beta\gamma}-\frac{1}{N+1}(\delta^{\alpha\gamma}
- \delta^{\alpha\delta}-\delta^{\beta\gamma}+\delta^{\beta\delta})
=P^{\alpha\gamma}P^{\beta\delta}-
P^{\alpha\delta}P^{\beta\gamma}\,.}[]
It is easy to show that
\eqn{Q^{\alpha\beta\gamma\delta}=4\lsp\Big(P^{\alpha\beta}P^{\alpha\gamma}
P^{\gamma\delta}-\frac{1}{N+1}(P^{\alpha\beta}P^{\alpha\delta}
+P^{\beta\gamma}P^{\gamma\delta})+\frac{1}{(N+1)^2}P^{\beta\delta}\Big)\,,
}[Qsimp]
and, using $\delta^{a\ne\beta}=1-\delta^{\alpha\beta}$, that
\eqna{R^{\alpha\beta\gamma\delta}&=-\frac{N+1}{2\llsp(N-1)}\lsp
Q^{\alpha\beta\gamma\delta}
+P^{\alpha\gamma}P^{\beta\delta}+P^{\alpha\delta}P^{\beta\gamma}
-\frac{2}{N}P^{\alpha\beta}P^{\gamma\delta}\,.}[Rsimp]
With \Qsimp and \Rsimp the crossing equation that arises from \tetraFour
can be easily worked out. We find
\eqn{\sum_{\text{S}_\tetra^+}\lambda_\cO^2\begin{pmatrix}
  0\\
  F^-_{\Delta,\lsp\ell}\\
  F^+_{\Delta,\lsp\ell}\\
  0
\end{pmatrix}+
\sum_{\text{V}_\tetra^+}\lambda_\cO^2\begin{pmatrix}
  0\\
  0\\
  -\frac{4}{N+1}F^+_{\Delta,\lsp\ell}\\
  F^-_{\Delta,\lsp\ell}
\end{pmatrix}+
\sum_{\text{Y}_\tetra^+}\!\lambda_\cO^2\begin{pmatrix}
  F^-_{\Delta,\lsp\ell}\\
  \frac{2\llsp (N-1)}{N}\lsp F^-_{\Delta,\lsp\ell}\\
  -\tfrac{(N+1)(N-2)}{N(N-1)}F^+_{\Delta,\lsp\ell}\\
  -\frac{N+1}{2\llsp(N-1)}F^-_{\Delta,\lsp\ell}
\end{pmatrix}+
\sum_{\text{A}_\tetra^{\lnsp-}}\lambda_\cO^2\begin{pmatrix}
  F^-_{\Delta,\lsp\ell}\\
  0\\
  F^+_{\Delta,\lsp\ell}\\
  0
\end{pmatrix}=\begin{pmatrix}
  0\\
  0\\
  0\\
  0
\end{pmatrix}.}[crEqTetra]
This crossing equation is equivalent to the one that recently appeared
in~\cite{Rong:2017cow}.\foot{To see this one needs to rescale, with
positive $N$-dependent factors as is necessary, the projectors defined by
those authors.}

Using the primitive invariant tensor $d_{ijkl}$ of~\cite{Osborn:2017ucf}
for the hypertetrahedral case and the relation
\eqn{d_{ijmn}d_{mnkl}=\frac{N(N-2)}{(N+1)(N+2)^2}\Big(
\delta_{ik}\delta_{jl}+\delta_{il}\delta_{jk}-\frac{2}{N}\delta_{ij}
\delta_{kl}\Big)
+\frac{N^2-3N-2}{(N+1)(N+2)}\lsp d_{ijkl}\,,}[]
we can find the projectors
\eqna{P_{ijkl}^{(1)}&=\frac{1}{N}\delta_{ij}\delta_{kl}\,,\\
  P_{ijkl}^{(2)}&=\frac{N+1}{N-1}\lsp d_{ijkl}+\frac{N}{(N-1)(N+2)}\Big(
\delta_{ik}\delta_{jl}+\delta_{il}\delta_{jk}-\frac{2}{N}\delta_{ij}
\delta_{kl}\Big)\,,\\
P_{ijkl}^{(3)}&=-\frac{N+1}{N-1}\lsp d_{ijkl}
+\frac{(N-2)(N+1)}{2(N-1)(N+2)}\Big(
\delta_{ik}\delta_{jl}+\delta_{il}\delta_{jk}-\frac{2}{N}\delta_{ij}
\delta_{kl}\Big)\,,\\
P_{ijkl}^{(4)}&= -\tfrac12(\delta_{ik}\delta_{jl}-
\delta_{il}\delta_{jk})\,.
}[]
Just like in the hypercubic case these satisfy \projProp.

It is straightforward to verify that for $N=3$ \crEqTetra and \crEqCubic
are equivalent, with $\text{V}^+_\tetra\leftrightarrow\text{Y}^+_\square$
and $\text{Y}^+_\tetra\leftrightarrow\text{V}^+_\square$ .  This
equivalence is a consequence of the fact that for $N=3$ the vertices of the
tetrahedron also form diagonals of the cube~\cite{Zia:1975ha}. For other
$N$ the hypertetrahedral group is not a subgroup of the hypercubic group.
We should note here that our symmetry group is really $\cS_{N+1}\times
\mathbb{Z}_2$.  The $\mathbb{Z}_2$ gives a minus sign to all fields, and it
can be broken if we assume that $\phi^\gamma$ appears in the OPE
$\phi^\alpha\times\phi^\beta$. In this work we do not make this assumption.

Let us now mention some results that the $\veps$ expansion gives for
hypertetrahedral theories~\cite{Osborn:2017ucf}. For generic $N$ there are
two hypertetrahedral fixed points with
\eqna{\Delta_\phi^{\!T^{(1)}_N}&=\tfrac12(d-2)+\frac{(N+1)(N+7)}{108\,
(N+3)^2}\veps^2
+\frac{(N+1)(109 N^3+969 N^2+5463 N+7411)}{11664 (N+3)^4}\,\veps^3\,,\\
\Delta_\phi^{\!T^{(2)}_N}&=\tfrac12(d-2)+
\frac{(N-1)(N-2)(N^2-6N+11)}{108\,(N^2 -5N + 8 )^2}\, \veps^2\\
&\hspace{-0.8cm}+\frac{(N-1)(N-2)(109 N^6-1752 N^5+12336 N^4-48804 N^3
+114807 N^2-151944 N+88208)}{11664(N^2-5 N+8)^4}\,\veps^3\,.
}[]
These fixed points do not exist for all $N$. In the framework of the
$\veps$ expansion one can see that the hypertetrahedral fixed points
collide and disappear into the complex plane at some $N_-$, and again
reappear at some $N_+$. These are given by~\cite{Osborn:2017ucf}
\eqn{
N_{\pm} =  5 + 6 \,\veps  +  \tfrac{281}{32}\, \veps^2
- \tfrac{61}{8} \, \zeta_3 \,\veps^2
\pm \sqrt{24\, \veps  + \tfrac{289}{4} \, \veps^2 - 30 \, \zeta_3
\, \veps^2} \, ,
}[Npm]
where $\zeta_3$ is Ap\'ery's constant. For $ N_- < N <  N_+$ there are no
hypertetrahedral fixed points. If we brazenly plug in $\veps=1$ to \Npm we
find
\eqn{N_\pm\approx10.62\pm 7.76\,.}[NpmEps]

\begin{figure}[ht]
  \centering
  \includegraphics{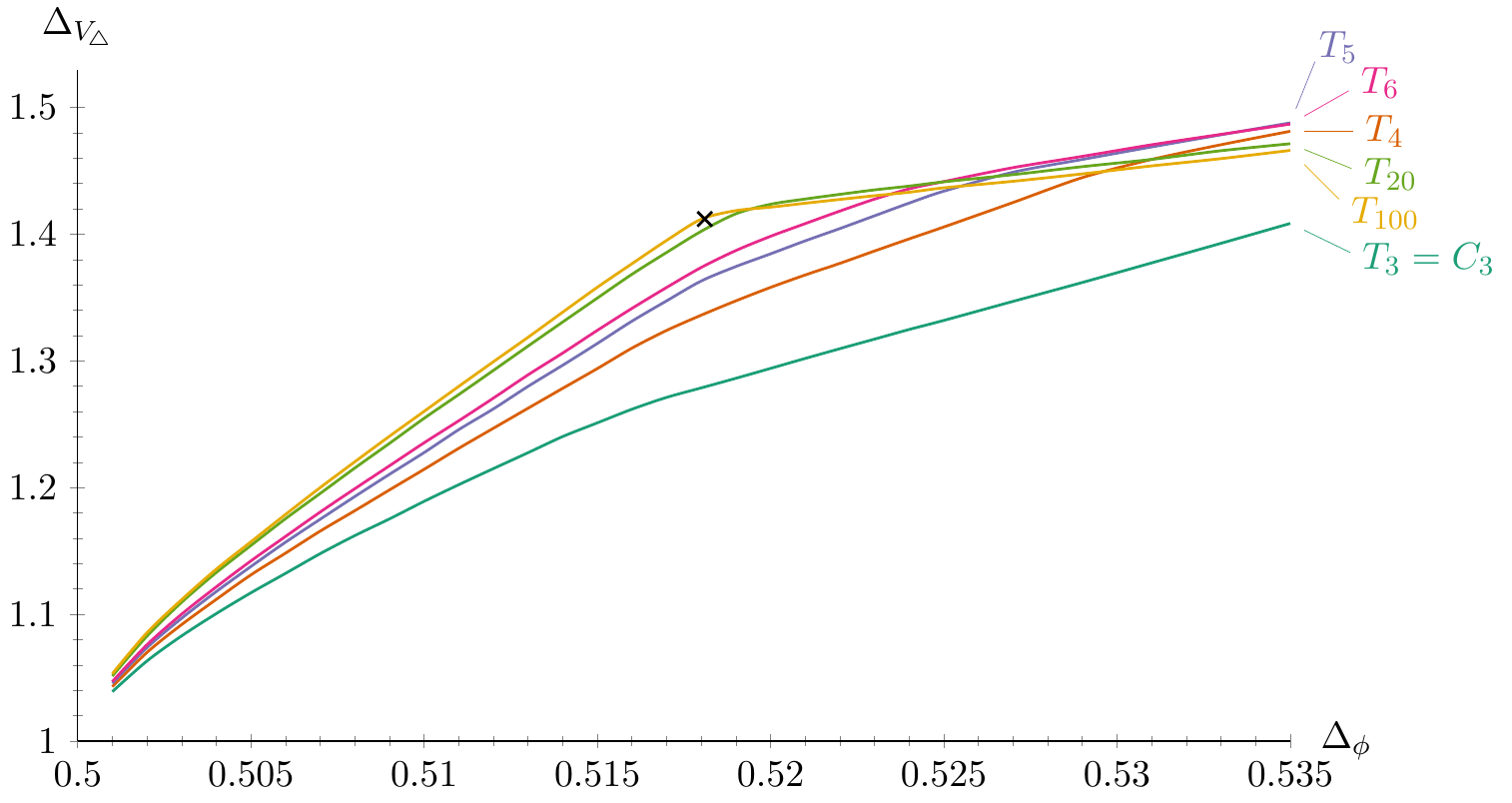}
  \caption{Upper bound on the dimension of the first scalar operator in the
  $\text{V}^+_\tetra$ sector of the $\phi^\alpha\times\phi^\beta$ OPE as a
  function of the dimension of $\phi$.  Areas above the curves are excluded
  in the corresponding theories. The x-marker indicates the position of the
  decoupled Ising theory.}
  \label{fig:Delta_V_tetra}
\end{figure}

\newsec{Bounds in hypertetrahedral theories}[boundsHyperT]
In the singlet sector the bound on $\Delta_{S_\tetra}$ is again saturated
by the $O(N)$ solution. A bound on the dimension of the first scalar
operator in the $\text{V}^+_\tetra$ sector is shown in
Fig.~\ref{fig:Delta_V_tetra}.
As we see, for low values of $N$ there is no kink in the bounds. The kink
develops as $N$ increases; this is consistent with the absence of CFTs with
hypertetrahedral symmetry for low values of $N$ as expected from \NpmEps.
We also see that at large $N$ the decoupled Ising theory is in the allowed
region and in fact saturates the bound.

\begin{figure}[H]
  \centering
  \includegraphics{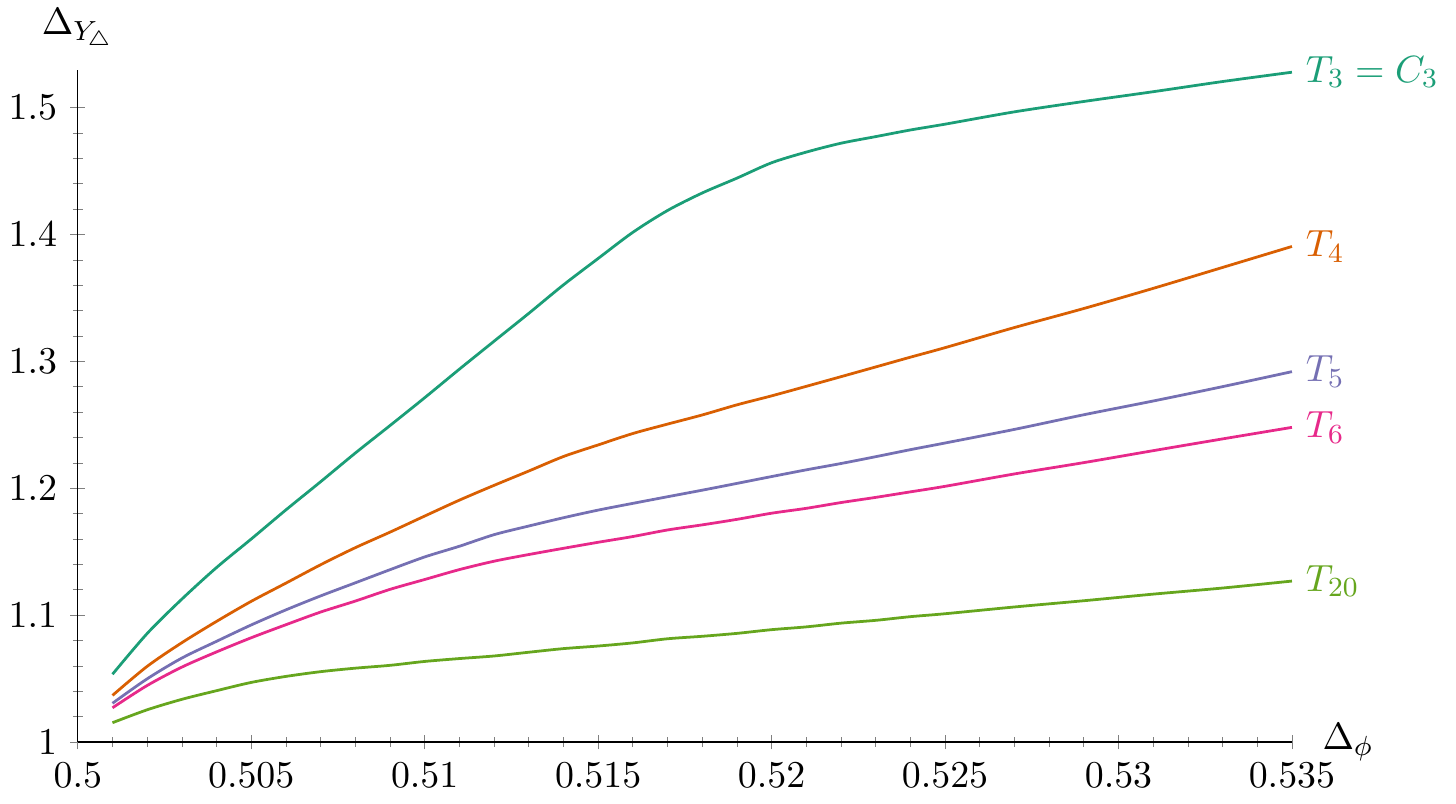}
  \caption{Upper bound on the dimension of the first scalar operator in the
  $\text{Y}^+_\tetra$ sector of the $\phi^\alpha\times\phi^\beta$ OPE as a
  function of the dimension of $\phi$.  Areas above the curves are excluded
  in the corresponding theories.}
  \label{fig:Delta_Y_tetra}
\end{figure}

Although for $N=3$ there is no kink in the bound of
Fig.~\ref{fig:Delta_V_tetra}, a kink appears in the $\text{Y}_\tetra^+$
sector as in Fig.~\ref{fig:Delta_Y_tetra}. Indeed, for $N=3$ we see a kink
in the dimension of $Y_\tetra\equiv V_\square$. This is the same as the
kink in the $C_3$ curve of Fig.~\ref{fig:Delta_V_cubic}, and so its
interpretation as a CFT is questionable. No kink is seen for other values
of $N$, however, consistently with our expectations from the $\veps$
expansion.

Although a precise determination of $N_{\pm}$ with the bootstrap of a
single correlator is perhaps not feasible, we expect $N_-\lesssim 4$ and
$N_+\lesssim15$ based on the presence of a kink in the bound of
$\Delta_{V_\tetra}$. We have looked at the spectrum of the solution that
saturates the bounds of Fig.~\ref{fig:Delta_V_tetra}, but we have not been
able to identify a feature that can serve as a conclusive indicator of the
existence of a CFT. It would be interesting to study this important problem
in a mixed-correlator setting.

\newsec{Conclusion}[conc]
In this paper we studied hypercubic and hypertetrahedral theories in $d=3$
with the non-perturbative numerical conformal bootstrap. We focused mainly
on the $N=3$ cubic theory due to its importance for phase transitions of
cubic magnets and perovskites. We found that the bound of the first scalar
operator in the $\text{V}_\square^+$ sector is saturated by a kink
solution, called $C_3^b$, with properties that cannot be reconciled with
the $\veps$ expansion.

There are (at least) three possibilities for the fate of $C_3^b$ (and
$C_N^b$):
\begin{enumerate}
  \item it corresponds to a hypercubic CFT that is inaccessible with the
    $\veps$ expansion,
  \item it corresponds to the hypercubic CFT found with the $\veps$
    expansion,
  \item it does not correspond to an actual CFT.
\end{enumerate}
Perhaps the most conservative possibility is the last one. The presence of
the kink would then be an artifact of the numerics. In any case, without
independent arguments at our disposal we cannot convincingly settle on the
correct interpretation of $C_N^b$.

\begin{figure}[H]
  \centering
  \includegraphics{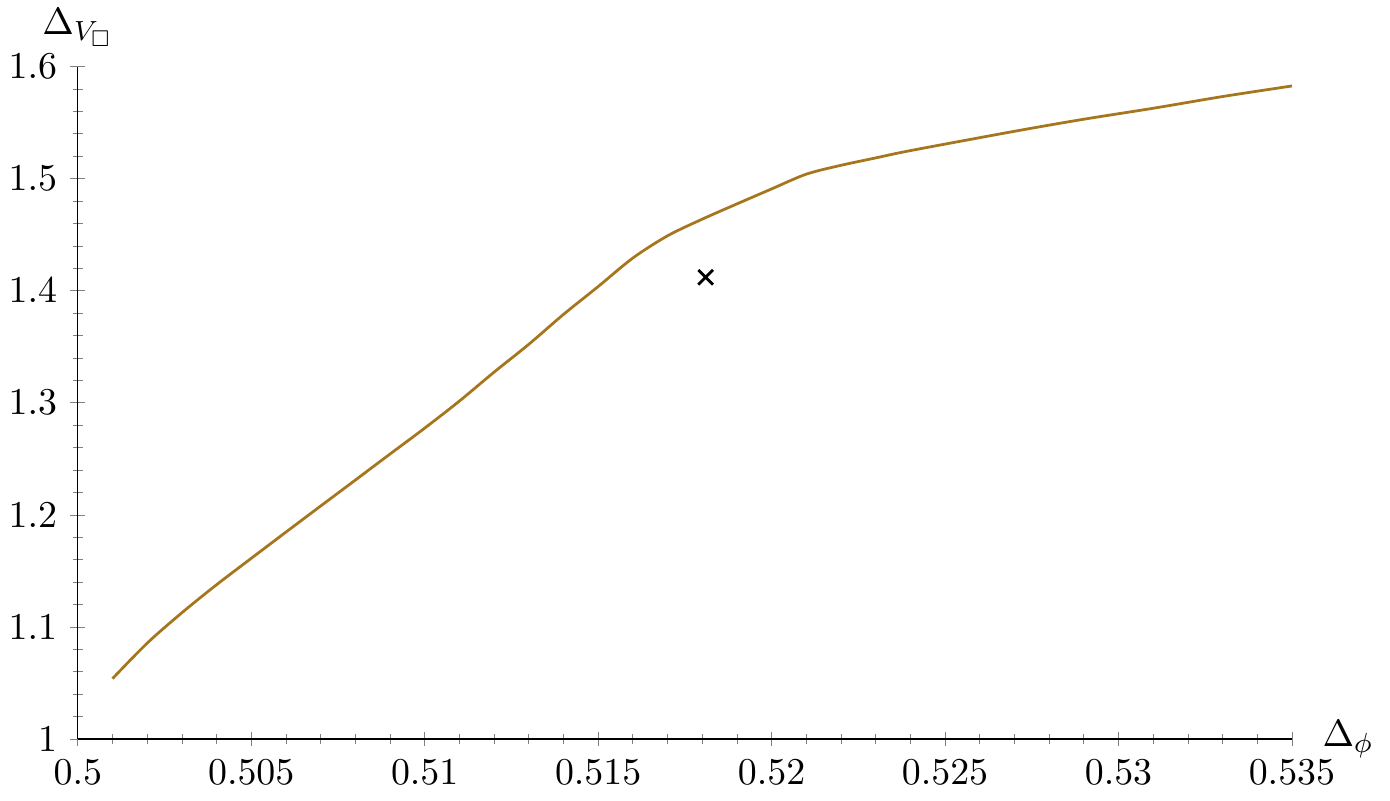}
  \caption{Upper bound on the dimension of the first $\text{V}_\square^+$
  scalar operator in the $\phi_i\times\phi_j$ OPE as a function of the
  dimension of $\phi$ for $N=2$. The area above the curve is excluded. The
  x-marker indicates the position of the decoupled Ising theory.}
  \label{fig:Delta_V_cubic_N2}
\end{figure}

According to the $\veps$ expansion for $N=2$ there is no fully-interacting
CFT other than the $O(2)$ model in $d=4-\veps$~\cite{Osborn:2017ucf}. The
only solution with ``cubic symmetry'' for $N=2$ is the decoupled Ising
model. We would thus expect the $N=2$ bound on $\Delta_{V_\square}$ to be
saturated at the Ising point.  This is, however, not what we find, as we
see in Fig.~\ref{fig:Delta_V_cubic_N2}. The bound is in fact weaker than
the $C_3$ bound of Fig.~\ref{fig:Delta_V_cubic}. The (smooth) kink observed
suggests that there may be a CFT not seen by the $\veps$ expansion, or that
a CFT with higher symmetry gets in the way. It could also be an artifact of
the numerics and correspond to no actual CFT.

Let us close by reminding the reader that when $N\to\infty$ the hypercubic
theory should go the constrained Ising model~\cite{Fisher:1968zzb,
Emery:1975zz, Aharony:1976}, in which the first scalar singlet has scaling
dimension $3-\Delta_\epsilon\approx1.5874$. This is a non-perturbative
result. We have checked that this is not what happens for $C_\infty^b$,
i.e.\ for the solution obtained at the kink of the bound on the dimension
of $V_\square$ at large $N$. However, in the
$\Delta_{V_\square}\text{-}\Delta_\phi$ slice of parameter space there are
two CFTs at the same exact position for $N\to\infty$. One is the decoupled
Ising theory, and the other is the constrained Ising one. In the former the
lowest-dimension singlet scalar has dimension
$\Delta_\epsilon\approx1.4126$, while in the latter it has dimension
$3-\Delta_\epsilon$. Therefore, the interpretation of the spectrum of the
$C_\infty^b$ solution in that case, solely derived from the solution at the
kink in the $\Delta_{V_\square}\text{-}\Delta_\phi$ slice, is unclear. It
is plausible that in order to derive meaningful results one needs to use
information from other sectors of operators, specifically those in which
the two CFTs do not have spectra of operators with the same scaling
dimensions.

\ack{I would like to thank Stefanos R.\ Kousvos and Theodore N.\ Tomaras
for collaboration in the initial stages of this project, and for many
helpful discussions. I also thank Connor Behan for help with
\texttt{PyCFTBoot}, and Kostas Siampos and Alessandro Vichi for very useful
discussions and comments. Finally, I am grateful to Amnon Aharony, Hugh
Osborn, David Poland, Slava Rychkov, and Alessandro Vichi for comments on
the manuscript. The numerical computations in this paper were run on the
LXPLUS cluster at CERN.}

\begin{appendices}

\newsec{An analysis of \texorpdfstring{\boldmath{$C_3^b$}}{C3b}
in \texorpdfstring{\boldmath{$d=3.8$}}{d=3.8}}[CIIIfrac]
In this appendix we study the $C_3^b$ solution in fractional spacetime
dimension, namely $d=3.8$. We expect our analysis not to be invalidated by
the fact that in such fractional dimensions unitarity is not expected to be
present~\cite{Hogervorst:2015akt}. For the Ising model a similar analysis
was performed in~\cite{El-Showk:2013nia}. Our aim is to explore the
properties of the $C_3^b$ solution, and compare with the $\veps$ expansion
at $\veps=0.2$, a value we expect is low enough for perturbative results to
be more trustworthy than in the $\veps=1$ case.\foot{We could also go to
smaller $\veps$, but the fact that the region where the kink is expected
moves towards the free theory makes the numerics slower. This is because of
the singular behavior of conformal blocks at the free theory.} Again,
$\Delta_{S_\square^\veps}$ and $\Delta_{S_{\mcirc}}$ are expected to be
very close according to the $\veps$ expansion. For the plots of this
appendix we use $\texttt{nmax}=9$, $\texttt{mmax}=6$, $\texttt{kmax}=40$,
$\texttt{cutoff}=10^{-15}$ in \texttt{PyCFTBoot}, and the same options and
parameters as before for \texttt{SDPB}. We find the bounds with a vertical
tolerance of $10^{-6}$.

\begin{figure}[H]
  \centering
  \includegraphics{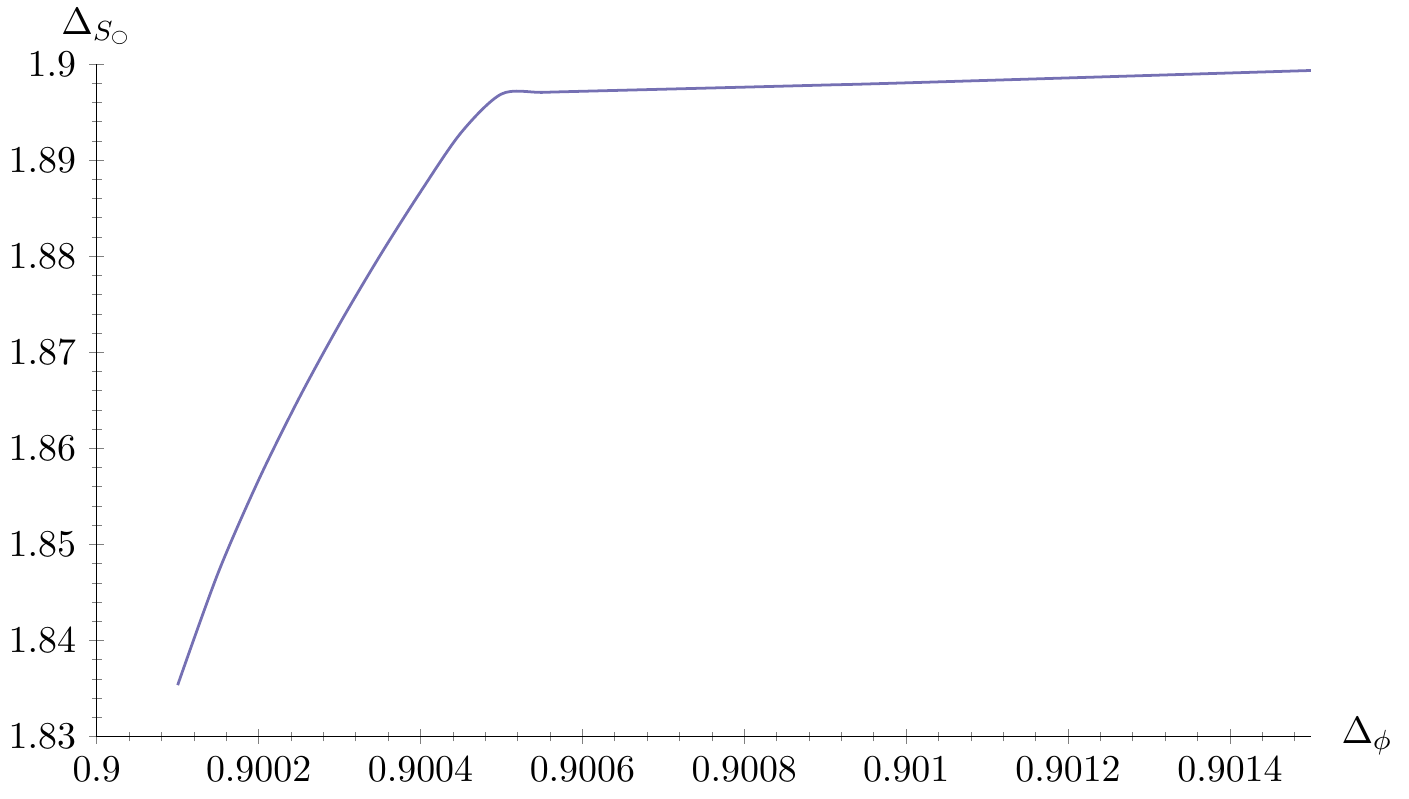}
  \caption{Upper bound on the dimension of the first singlet scalar in the
  $\phi_i\times\phi_j$ OPE as a function of the dimension of $\phi$ in the
  $O(3)$ model in $d=3.8$.} \label{fig:Delta_S_38_Heisenberg3}
\end{figure}

First, we obtain a bound on the first singlet scalar in the $O(3)$ model,
using the $O(N)$ crossing equation directly~\cite{Kos:2013tga}. The bound
is shown in Fig.~\ref{fig:Delta_S_38_Heisenberg3}, and it displays a strong
kink around $\Delta_\phi=0.9005$. If the $O(3)$ model lives on the kink,
then $\Delta_{S_{\mcirc}}\approx 1.8969$.

\begin{figure}[ht]
  \centering
  \includegraphics{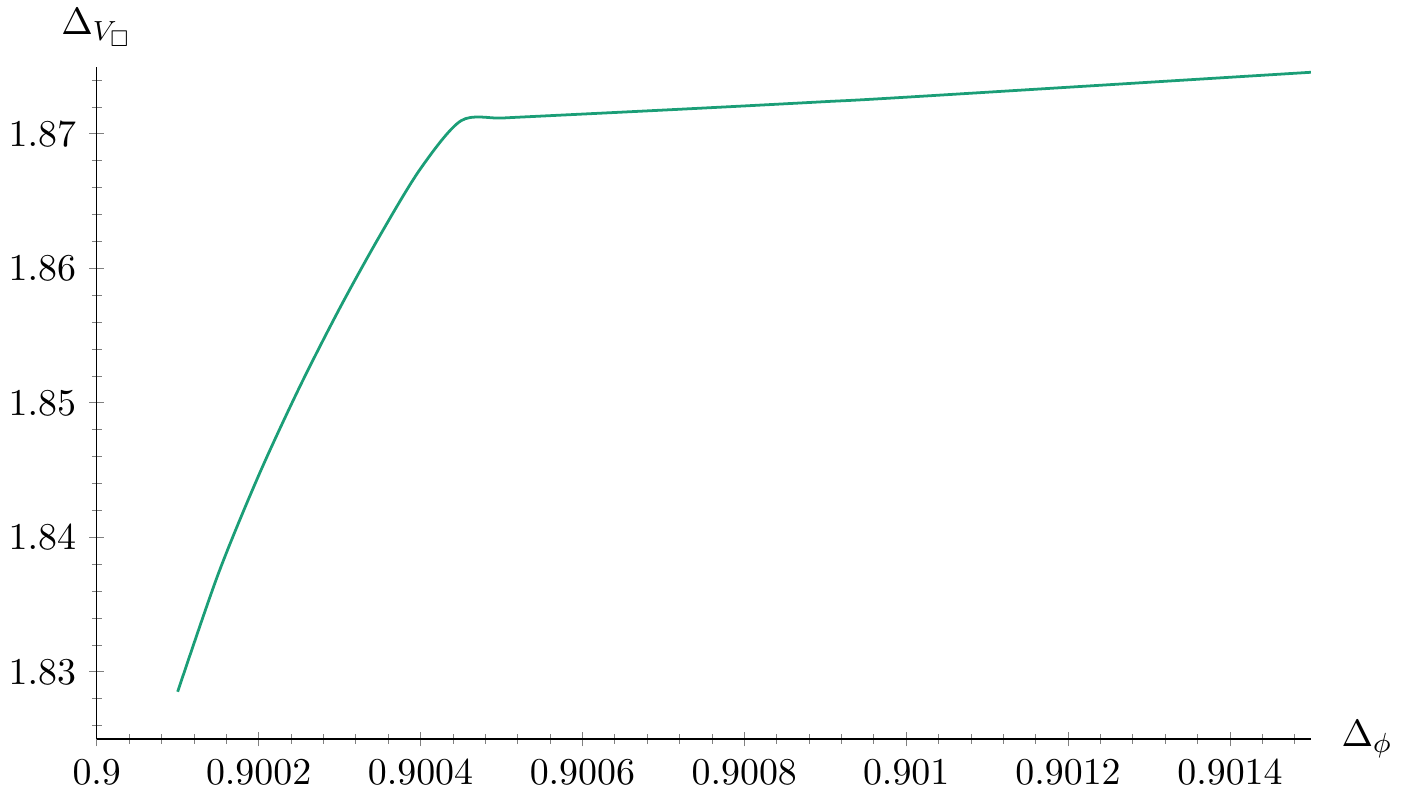}
  \caption{Upper bound on the dimension of the first $\text{V}_\square^+$
  scalar operator in the $\phi_i\times\phi_j$ OPE as a function of the
  dimension of $\phi$ in the $C_3$ theory in $d=3.8$.}
  \label{fig:Delta_V_38_cubic3}
\end{figure}

The bound on the scaling dimension of $V_{\square}$ is shown in
Fig.~\ref{fig:Delta_V_38_cubic3}. There is again a very sharp kink.  On the
bound at $\Delta_\phi\approx 0.90045$ we have
$\Delta_{V_\square}\approx1.8710$. At this point it is not easy to tell if
the kink is actually saturated by the decoupled Ising theory, for in the
latter case the dimension of $V_{\square}$ would be equal to the dimension
of the $\epsilon$ operator in the Ising model, which in $d=3.8$ is very
close to 1.87.  This can be seen both with the $\veps$
expansion~\cite{Kleinert:1991rg}, and by looking at the relevant kink
in~\cite{El-Showk:2013nia}. To resolve this we look at the
spectrum.\foot{We have also obtained the bound on $\Delta_{V_\square}$ in
$C_{20}$ in $d=3.8$, and at the corresponding kink $\Delta_{V_\square^b}$
is slightly lower than that in the $C_3^b$ solution of
Fig.~\ref{fig:Delta_V_38_cubic3}.  This indicates that we have not
saturated the bound with the decoupled Ising theory in
Fig.~\ref{fig:Delta_V_38_cubic3}.}

\begin{figure}[H]
  \centering
  \includegraphics{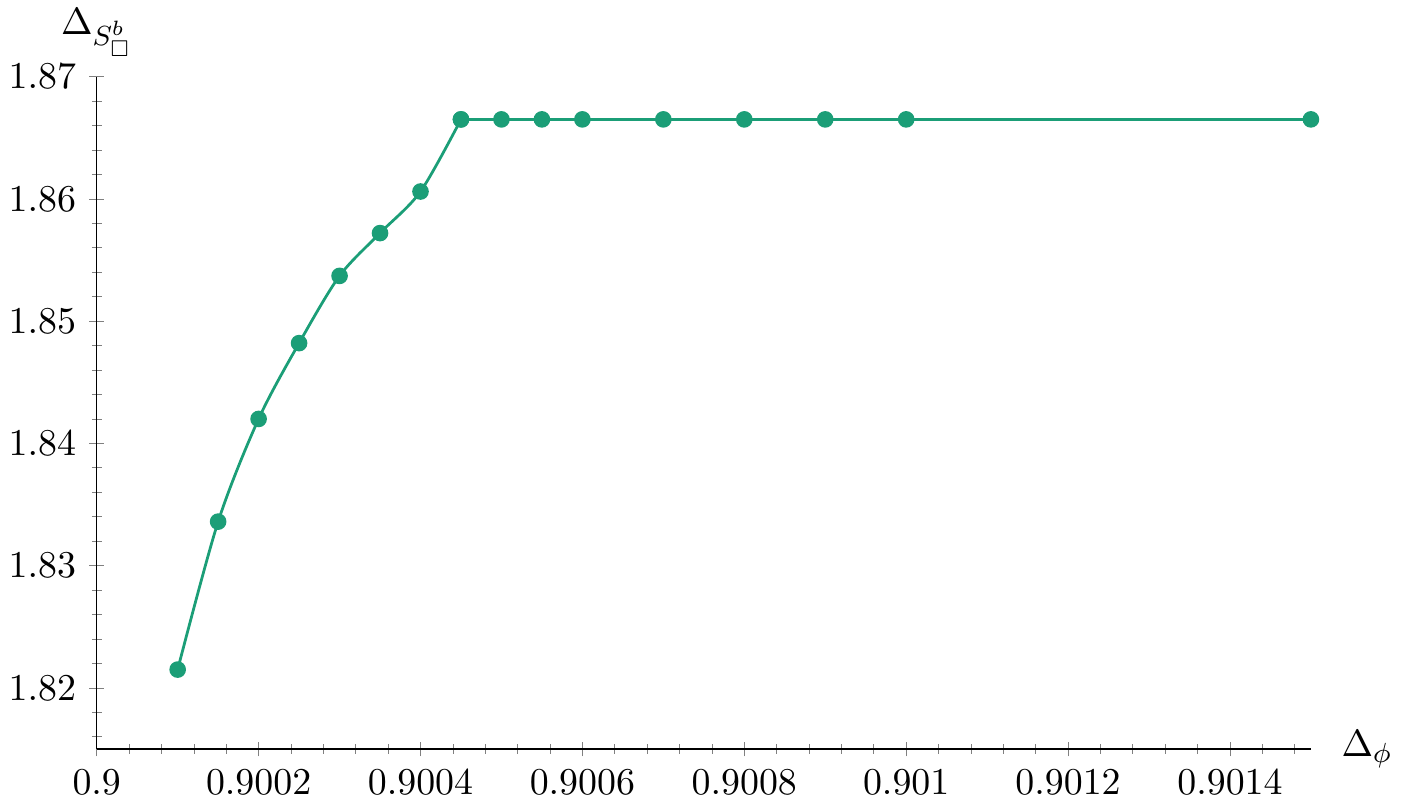}
  \caption{The dimension of $S_\square^b$ at the $C_3^b$ solution in
    $d=3.8$, i.e.\ assuming that $\Delta_{V_\square}$ is equal to the bound
    of Fig.~\ref{fig:Delta_V_38_cubic3}. The markers indicate the points at
    which we have computed the spectrum.}
  \label{fig:Delta_S_from_spectrum_38_cubic3}
\end{figure}

\begin{figure}[H]
  \centering
  \includegraphics{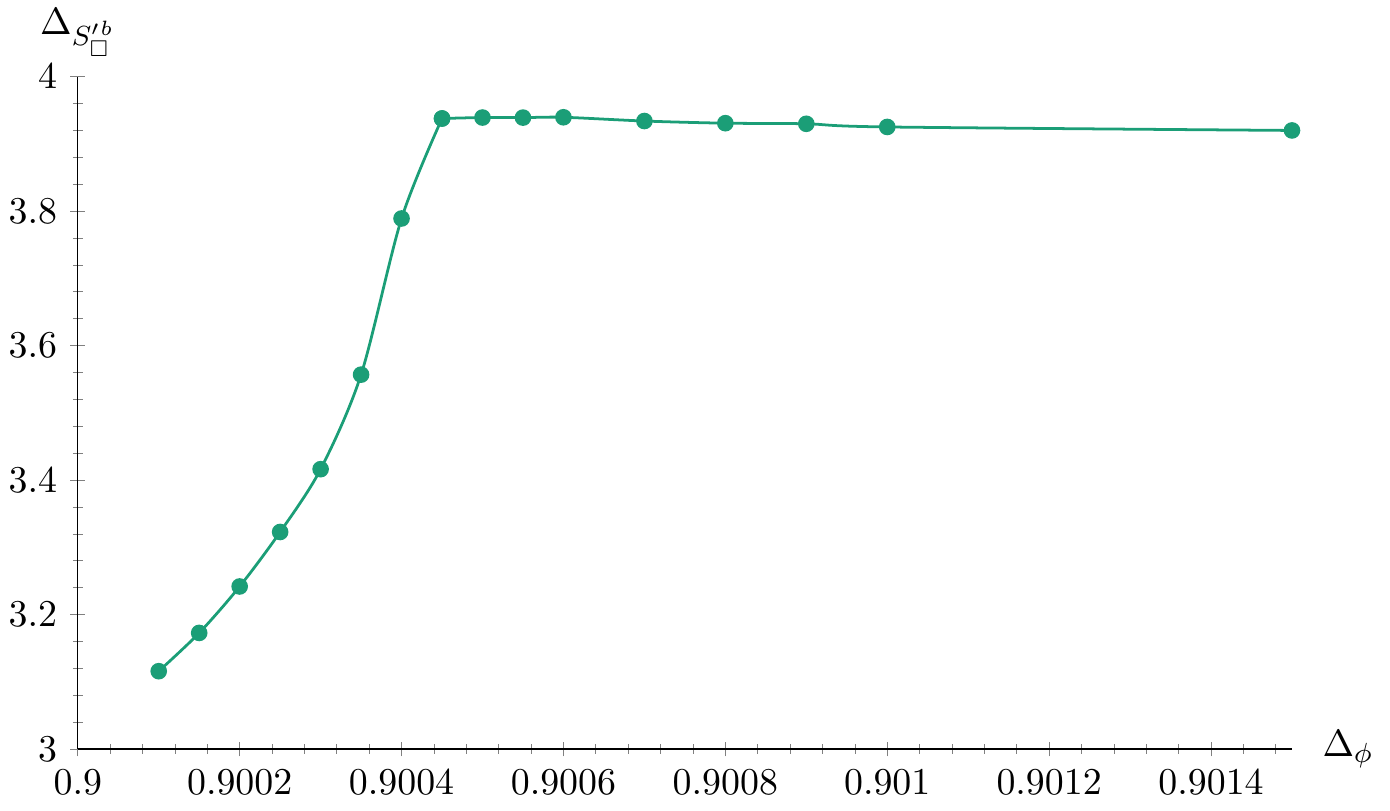}
  \caption{The dimension of $S_\square^{\prime\llsp b}$ at the $C_3^b$
    solution in $d=3.8$, i.e.\ assuming that $\Delta_{V_\square}$ is equal
    to the bound of Fig.~\ref{fig:Delta_V_38_cubic3}. The markers indicate
    the points at which we have computed the spectrum.}
  \label{fig:Delta_Sp_from_spectrum_38_cubic3}
\end{figure}

The scaling dimensions of the first and second scalar singlet of $C_3^b$
are shown in Figs.~\ref{fig:Delta_S_from_spectrum_38_cubic3}
and~\ref{fig:Delta_Sp_from_spectrum_38_cubic3}, respectively. At the kink
we have $\Delta_{S_\square^b}\approx1.8665$. Although close to
$\Delta_{V_\square^b}\approx1.8710$, the two dimensions are different which
gives us confidence that we are not in the decoupled Ising theory. For the
second scalar singlet we have $\Delta_{S_\square^{\prime\llsp b}}>3.8$,
which means that the $C_3^b$ solution is critical in $d=3.8$.

Finally we can compare with $\Delta_{S_{\mcirc}}\approx1.8969$ in $d=3.8$.
For $\veps=0.2$ we have $\Delta_{S_{\mcirc}}-\Delta_{S_{\square}^b}\approx
0.03$. The results for $\Delta_{S_{\mcirc}}$ and $\Delta_{S_{\square}^b}$
are significantly closer compared to the $\veps=1$ case considered in the
main text, where $\Delta_{S_{\mcirc}}-\Delta_{S_{\square}^b}\approx
0.2667$. Nevertheless, we would expect a much smaller
$\Delta_{S_{\mcirc}}-\Delta_{S_{\square}^b}$ if the $C_3^b$ solution at the
kink were to become the $C_3^\veps$ CFT found with the $\veps$ expansion at
$\veps=0.2$. Using the five-loop results of~\cite{Kleinert:1991rg,
Kleinert:1994td} we find, using a simple Pad\'e[3,\lsp 4] approximant,
$\Delta_{S_{\mcirc}}-\Delta_{S_{\square}^{\veps=0.2}}\approx -2.5\times
10^{-4}$. The negative sign here seems to be at odds with the bootstrap
results, but, given the smallness of
$|\Delta_{S_{\mcirc}}-\Delta_{S_{\square}^{\veps=0.2}}|$, we expect this
issue to be resolved by considering higher-order corrections and/or more
robust resummation techniques. With the same Pad\'e[3,\lsp 4] approximant
we find $\Delta_{S_{\mcirc}}-\Delta_{S_{\square}^{\veps=1}}\approx 0.01$.
At $\veps=1$ more advanced resummation techniques give
$\Delta_{S_{\mcirc}}-\Delta_{S_{\square}^{\veps=1}}\approx 6\times
10^{-4}$~\cite{Carmona:1999rm, Pelissetto:2000ek}. From this analysis we
may then conclude that it is not likely that the $C_3^b$ solution at the
kink is related to the $C_3^\veps$ cubic CFT. However, given that there is
a sharp kink in Fig.~\ref{fig:Delta_V_38_cubic3} we may speculate that the
$C_3^b$ solution is in fact accessible with the $\veps$ expansion. We hope
to return to these issues in the future.

\end{appendices}

\bibliography{bootstrapping_hypercubic_hypertetrahedral}

\end{document}